\documentclass[aps,prd,twocolumn,preprintnumbers,floatfix,nofootinbib]{revtex4}

\usepackage{multirow, graphicx,amssymb,url,mathrsfs,amsmath}
\usepackage{eucal,wrapfig,boxedminipage,setspace,subfigure,epsfig}
\usepackage{amsxtra,amstext,latexsym,dsfont,amsfonts}
\usepackage{makeidx,pst-grad,pst-plot}
\usepackage{color}
\usepackage[colorlinks,pagebackref,hyperindex]{hyperref}
    \hypersetup
    {
        colorlinks,%
        citecolor=red,%
        linkcolor=blue,%
        urlcolor=black,%
    }

\def\a  {\alpha}                
       \def\d  {\delta}        
\def\e  {\epsilon}

 \newcommand{\call}{\mbox{${\cal L}$}}

\def\IR{{\hbox{{\rm I}\kern-.2em\hbox{\rm R}}}}
\def\IB{{\hbox{{\rm I}\kern-.2em\hbox{\rm B}}}}
\def\IN{{\hbox{{\rm I}\kern-.2em\hbox{\rm N}}}}
\def\IC{\,\,{\hbox{{\rm I}\kern-.59em\hbox{\bf C}}}}
\def\IZ{{\hbox{{\rm Z}\kern-.4em\hbox{\rm Z}}}}
\def\IP{{\hbox{{\rm I}\kern-.2em\hbox{\rm P}}}}
\def\IH{{\hbox{{\rm I}\kern-.4em\hbox{\rm H}}}}
\def\ID{{\hbox{{\rm I}\kern-.2em\hbox{\rm D}}}}

\def\be{\begin{equation}}
\def\ee{\end{equation}}
\def\ba{\begin{eqnarray}}
\def\ea{\end{eqnarray}}

\def\ra{\rightarrow}  
\def\lra{\Longrightarrow}

\def\del{\partial}

\newcommand{\brac}[1]{\langle #1 \rangle}

\def\nn{\nonumber}
\def\ea{{\it et al}. }

\newcommand{\ie}{{\it i.e.}}

\newcommand{\wt}{\widetilde}

\newcommand{\beq}{\begin{equation}}
\newcommand{\eeq}{\end{equation}}
\newcommand{\bea}{\begin{eqnarray}}
\newcommand{\eea}{\end{eqnarray}}

\newcommand{\tw}{{{\widetilde w}}}

\newcommand{\trho}{{{\widetilde \rho}}}
\newcommand{\td}{{{\widetilde d}}}
\newcommand{\tL}{{{\widetilde L}}}
\newcommand{\tc}{{{\widetilde c}}}
\newcommand{\tm}{{{\widetilde m}}}
\newcommand{\tK}{{{\widetilde K}}}
\newcommand{\tmu}{{{\widetilde \mu}}}
\newcommand{\tF}{{{\widetilde F}}}
\newcommand{\tOmega}{{{\widetilde \Omega}}}

\begin{document}

\newcommand\sect[1]{\emph{#1}---}

\preprint{
\begin{minipage}[t]{3in}
\begin{flushright} SHEP-10-01
\\[30pt]
\hphantom{.}
\end{flushright}
\end{minipage}
}

\title{Holographic Description of the Phase Diagram of \\
a Chiral Symmetry Breaking Gauge Theory}

\author{Nick Evans}
\email{evans@soton.ac.uk}
\author{Astrid Gebauer}
\email{ag806@soton.ac.uk}
\author{Keun-Young Kim}
\email{k.kim@soton.ac.uk}
\author{Maria Magou}
\email{mm21g08@soton.ac.uk}

\affiliation{ School of Physics and Astronomy, University of
Southampton, Southampton, SO17 1BJ, UK \\ % \vspace{0.2cm}
}

\begin{abstract}

\noindent The large $N$ ${\cal N}$=4 gauge
theory with quenched ${\cal N}$=2 quark matter in the presence of
a magnetic field displays chiral symmetry breaking. We study the
temperature and chemical potential dependence of this theory using
its gravity dual (based on the D3/D7 brane system).  With massless
quarks, at zero chemical potential, the theory displays a first
order thermal transition where chiral symmetry is restored and
simultaneously the mesons of the theory melt. At zero temperature,
these transitions with chemical potential are second order and
occur at different chemical potential values. Between the three
there are two tri-critical points, the positions of which we
identify. At finite quark mass the second order transition  for
chiral symmetry becomes a cross over and there is a critical point
at the end of the first order transition, while the meson melting
transition remains similar to the massless quark case. We track
the movement of the critical points as the mass is raised relative
to the magnetic field.
%- we track {\bf their positions} in the plane as a function of quark mass.
%In addition there is a first order meson melting transition
%in the plane which joins the critical line above at a second
%critical point.

\noindent

\end{abstract}

\maketitle

\section{Introduction}

The phase diagram in the temperature chemical potential (or density) plane
is a matter of great interest in both QCD and
more widely in gauge theory
\cite{Rajagopal:2000wf,Stephanov:2007fk,Owe}. In QCD there is
believed to be a transition from a confining phase with chiral
symmetry breaking at low temperature and density to a phase with
deconfinement and no chiral symmetry breaking at high temperature.
In the standard theoretical picture for QCD with massless quarks,
the transition is first order for low temperature but growing
density, whilst second order at low density and growing
temperature. The second order transition becomes a cross over at
finite quark mass. There is a (tri-)critical point where the first
order transition mutates into the (second order) cross over
transition. In fact though there could still be room in QCD for a
more exotic phase diagram~\cite{Owe} as we will discuss in the
context of our results in our final section.

In this paper we will present a precise holographic
\cite{Malda,Witten:1998qj,Gubser:1998bc} determination of the
phase diagram in the temperature chemical potential plane for a
gauge theory that displays many of the features of the QCD
diagram, although the precise details differ. A pictorial
comparison of our theory to QCD can be made by comparing Fig
\ref{Tvsmu} to Fig \ref{QCD}.

The theory we will consider is the large $N$ ${\cal N}$=4 gauge
theory with quenched ${\cal N}$=2 quark matter
\cite{Karch,Polchinski,Bertolini:2001qa,Mateos} which has been
widely studied~\cite{Erdmenger:2007cm}. An immediate difference
between the ${\cal N}$=4 glue theory and QCD is that the thermal
phase transition to a deconfined phase occurs for infinitesimal
temperature since the massless theory is conformal
\cite{Witten:1998qj}. Essentially the entire temperature chemical potential
phase diagram of our theory is therefore characterized by strongly
coupled deconfined glue.

The quark physics is more subtle though - the phase diagram in the
temperature chemical potential (density) plane for the ${\cal
N}=2$ quark matter has been studied in
\cite{Myers1,Myers3,Karch1,Sin1,Sin2,Ghoroku}. When the quark mass
is zero the theory is conformal and the origin of the phase
diagram is a special point with confined matter. Immediately away
from that point, in either temperature or chemical potential, a
first order transition moves the theory to a deconfined theory
(the mesons melt \cite{Peeters:2006iu,Hoyos:2006gb,Erdmenger0}).

When a quark mass is present in the ${\cal N}$=2 theory the meson
melting transition occurs away from the origin. This transition
has been reported as first order with a second order transition
point where the first order transition line touches the $T =0$
chemical potential axis~\cite{Myers3,Karch1} (in the grand
canonical ensemble). Interestingly there is a phase transition
line in the temperature versus density plane (in the canonical
ensemble) in which the quark condensate jumps~\cite{Sin1,Myers1}. This
area of the phase diagram is intrinsically unstable though and not
realizable by imposing any chemical potential~\cite{Myers3}.

The crucial ingredient we will add to the theory is chiral
symmetry breaking which will also bring the theory closer in
spirit to QCD. As shown in \cite{Johnson1,Erdmenger1,Zayakin,
Johnson2} the ${\cal N}=2$ theory in the presence of a magnetic
field displays chiral symmetry breaking through the generation of
a quark anti-quark condensate. At zero density the finite
temperature behaviour has been studied \cite{Johnson1,Erdmenger1}
and there is a first order transition from a chiral symmetry
broken phase at low temperature to a chiral symmetry restored
phase at high temperature. In this paper we will include chemical
potential as well to map out the full phase diagram in the
temperature chemical potential plane. We will find a chiral
symmetry restoration phase transition, which is first order for
low density and second order for low temperature - there is a
critical point where these transitions meet. This physics is in
addition to a meson melting transition which is first order at
large temperature but apparently second order at low temperature.
This latter region of transition is interesting because it is
associated with a discontinuous jump from an embedding off the
black hole to one that ends on it and it looks naively first
order. However, when we plot any available order parameter in the
boundary theory it appears second order.

 We will also track the movement of these transition lines and
critical points as the quark mass rises relative to the magnetic
field. The infinite mass limit corresponds to the pure ${\cal
N}$=2 theory without magnetic field~\cite{Myers3,Sin2}. 
The second order chiral
symmetry restoration transition becomes a cross over the moment a
mass is introduced. The first order transition structure though
remains, even to the infinite mass limit, with two critical
points: one is the end point of the first order transition and the
other is the the end point of the second order meson melting
transition. This structure was not reported in the results in
\cite{Myers3,Sin2}\footnote{ The existence of 
two critical points is related with the existence of 
the black hole to black hole transition. It  
is actually just visible in Fig 2c of \cite{Sin2} but 
the authors had not probed it in detail previously. 
After discussion of our results with the authors of \cite{Sin2}, 
they have refined their computations and confirmed our results.} 
but this is not
surprising since the structure, in that limit, is on a very fine
scale. We have only found it by following the evolution of the
larger structure present at low quark mass with a magnetic field.
In addition we present evidence to suggest the parameter space
with a second order meson melting transition extends away from
just the $T=0$ axis, again, even in the infinite mass limit. We
have confirmed these results in the strict $B=0$ limit also.

The theory we study may appear to be a rather vague relative of
QCD with magnetic field induced chiral symmetry breaking. On the
other hand it is a theory of strongly coupled glue with the
magnetic field inducing conformal symmetry breaking in the same
fashion as $\Lambda_{\rm QCD}$ in QCD. In fact the magnetic field
case in the basic ${\cal N}$=4 dual is the cleanest known example
of chiral symmetry breaking in a holographic environment. Other
deformations of the ${\cal N}$=4 gauge theory typically lead to an
ill-understood IR singular hard wall - see for example
\cite{Gubser:1999pk,Girardello:1999bd}. The magnetic field case
provides a smooth IR wall where we have more control but the
results are likely to be the same in those more complex cases. We
can hope to learn some lessons for a wider class of gauge
theories.

\section{The holographic description}

The ${\cal N}$=4 gauge theory at finite temperature has a
holographic description in terms of an AdS$_5$ black hole geometry
(with $N$ D3 branes at its
core)\cite{Malda,Witten:1998qj,Gubser:1998bc}. The geometry is
\begin{eqnarray}
  ds^2 = \frac{r^2}{R^2}(-f dt^2 + d\vec{x}^2) + \frac{R^2}{r^2 f} dr^2
  + R^2 d\Omega_5^2 \ ,
\end{eqnarray}
where $R^4=4 \pi g_s N \alpha^{'2}$ and
\begin{eqnarray}
  f := 1-\frac{r_H^4}{r^4} \ , \qquad r_H := \pi R^2 T \ .
\end{eqnarray}
Here $r_H$ is the position of the black hole horizon which is
related to the temperature $T$.

We will find it useful to make the coordinate transformation
\begin{eqnarray}
  \frac{dr^2}{r^2f} \equiv \frac{dw^2}{w^2}
   \ \lra \  w := \sqrt{r^2 + \sqrt{r^4 - r_H^4}}\ , \label{rtow}
\end{eqnarray}
with $w_H = r_H$. This change makes the presence of a flat 6-plane
perpendicular to the horizon manifest. We will then write the
coordinates in that plane as $\rho$ and $L$ according to
%{\it Note:} Another convention used in the literature
%is $\sqrt{2} w := \sqrt{r^2 + \sqrt{r^4 - r_H^4}}$
%with $\sqrt{2} w_H = r_H $. In this case $w=r$ when $T=0$ so reduced
%to zero temperature coordinate. However in this paper we adopt (\ref{rtow})
%since many works we are referring \cite{Myers1, Johnson1, Johnson2, Erdmenger1} are using (\ref{rtow}).
\begin{eqnarray}
  w = \sqrt{\rho^2 + L^2}\ ,  \quad \rho := w\sin\theta \ .
  \quad L := w\cos\theta \ ,
\end{eqnarray}

The metric is then
\begin{eqnarray}
   ds^2 &=& \frac{w^2}{R^2}(- g_t dt^2 + g_x d\vec{x}^2) \nn \\
        & & + \frac{R^2}{w^2} (d\rho^2 + \rho^2 d\Omega_3^2
         + dL^2 + L^2 d\Omega_1^2) \ ,
\end{eqnarray}
where
\begin{eqnarray}
g_t := \frac{(w^4 - w_H^4)^2}{2 w^4 (w^4+w_H^4)}\ ,  \qquad
g_x  := \frac{w^4 + w_H^4}{ 2 w^4} \ .
\end{eqnarray}

\subsection{Quarks/D7 brane probes}

Quenched ($N_f \ll N$) ${\cal N}$=2 quark superfields can be
included in the ${\cal N}$=4 gauge theory through probe D7 branes
in the geometry\cite{Karch,Polchinski,Bertolini:2001qa,Mateos}.
The D3-D7 strings are the quarks. D7-D7 strings holographically
describe mesonic operators and their sources. The D7 probe can be
described by its DBI action
\beq S_{DBI} = - T_{D7} \int d^8\xi \sqrt{- {\rm det} (P[G]_{ab} +
2 \pi \alpha' F_{ab})} \ , \eeq
where $P[G]_{ab}$ is the pullback of the metric and $F_{ab}$ is
the gauge field living on the D7 world volume. We will use
$F_{ab}$ to introduce a constant magnetic field (eg $F_{12} = -
F_{21} = B$)~\cite{Johnson1} and a chemical potential associated with baryon
number $A_t(\rho) \neq 0$~\cite{Myers1,Kim}.

We embed the D7 brane in the $\rho$ and $ \Omega_3$ directions of
the metric but to allow all possible embeddings must include a
profile $L(\rho)$ at constant $\Omega_1$. The full DBI action we
will consider is then
\begin{eqnarray}
  S = \int d\xi^8 \call(\rho)
    = \left(\int_{S^3} \e_3 \int dtd\vec{x} \right) \int d\rho \
  \call(\rho) \ ,
\end{eqnarray}
where $\e_3$ is a volume element on the 3-sphere and
\begin{eqnarray}
  \call &:=& -N_f T_{D7} \frac{\rho^3}{4}\left(1-\frac{w_H^4}{w^4}\right)
  \nn \\
  &&\times \sqrt{\left(1+(\partial_\rho L)^2
   - \frac{ 2 w^4 (w^4+w_H^4)}{(w^4 - w_H^4)^2} (2\pi\a' \partial_\rho A_t)^2 \right)}
   \nn \\
  &&\times \sqrt{\left(\left(1+\frac{w_H^4}{w^4}\right)^2 + \frac{4 R^4}{w^4}B^2 \right)} \ . \label{OriginalAction}
\end{eqnarray}
Since the action is independent of $A_t$, there is a conserved
quantity $d$ $\left(:= \frac{\delta S}{\delta F_{\rho t}}\right)$
and we can use the Legendre transformed action
\begin{eqnarray}
  \wt{S} = S - \int d\xi^8 F_{\rho t} \frac{\delta S}{\delta F_{\rho t}}
         =  \left(\int_{S^3} \e_3 \int dtd\vec{x} \right) \int d\rho \
  \wt{\call}(\rho) \ , \nn \\ \label{LegendreAction}
\end{eqnarray}
where
\begin{eqnarray}
  &&\wt{\call} := - N_f T_{D7} \frac{(w^4-w_H^4)}{4 w^4}
  \sqrt{K (1+(\partial_\rho L)^2)} \label{Hamiltonian} \\
      && K  :=  \left(\frac{w^4+w_H^4}{w^4}\right)^2 \rho^6
    + \frac{4  R^4 B^2}{w^4} \rho^6 \nn \\
    && \qquad + \frac{8 w^{4}}{(w^4+w_H^4)}
    \frac{d^2}{(N_f T_{D7} 2\pi\a')^2} \ .
\end{eqnarray}
To simplify the analysis we note that we can use the magnetic
field value as the intrinsic scale of conformal symmetry breaking
in the theory - that is we can rescale all quantities in
(\ref{Hamiltonian})  by $B$ to give
\begin{eqnarray} \label{rescale}
  &&\wt \call = -N_f T_{D7} (R\sqrt{B})^4 \
        \frac{\tw^4 - \tw_H^4}{\tw^4}\sqrt{\tK(1+\tL'^2)}\ ,
        \quad \label{tH}\\
   &&\tK = \left(\frac{\tw^4+\tw_H^4}{\tw^4}\right)^2 \trho^6
                +\frac{1}{\tw^4} \trho^6
              + \frac{\tw^4}{(\tw^4+\tw^4_H)} \td^2 \ , \quad
\end{eqnarray}
where the dimensionless variables are defined as
\begin{eqnarray}
   && (\tw , \tL, \trho, \td) \nn \\
   && := \left(\frac{w}{R\sqrt{2B}},
        \frac{L}{R\sqrt{2B}} ,
       \frac{\rho}{R\sqrt{2B}},
       \frac{d}{(R \sqrt{B})^3 N_f T_{D7} 2\pi \a'}\right) \ . \nn \\
\end{eqnarray}

In all cases the embeddings become flat at large $\rho$  taking
the form
\begin{eqnarray}
  \tL(\trho) \sim \tm + \frac{\tc}{\trho^2}\ ,
  \tm = \frac{2\pi \a' m_q}{R\sqrt{2B}} \ ,
  \tc = \brac{\bar{q}q} \frac{(2\pi \a')^3}{(R\sqrt{2B})^3} \ . \ \label{scaled}
\end{eqnarray}
In the absence of temperature, magnetic field and density
 the regular embeddings are simply $L(\trho)= \tm$, which is the
minimum length of a D3-D7 string, allowing us to identify it with
the quark mass as shown. $\tilde{c}$ should then be identified
with the quark condensate with the relation shown.

We will classify the D7 brane embeddings  by their small $\trho$
behavior. If the D7 brane touches the black hole horizon, we call
it a black hole embedding, otherwise, we call it a Minkowski
embedding.  We have used Mathematica to solve the equations of
motion for the D7 embeddings resulting from (\ref{rescale}).
Typically in what follows, we numerically shoot out from the black
hole horizon (for black hole embeddings) or the $\trho=0$ axis
(for Minkowski embeddings) with Neumann boundary condition for a
given $\td$. Then by fitting the embedding function with
(\ref{scaled}) at large $\trho$ we can read off $\tm$ and $\tc$.

\subsection{Thermodynamic potentials}

The Hamilton's equations from (\ref{LegendreAction}) are
$\del_\rho d = \frac{\d\wt{S}}{\d A_t} $ and
$\del_\rho A_t = - \frac{\d\wt{S}}{\d d} $. The first simply means
that $d$ is the conserved quantity. The second reads as
\begin{eqnarray} \label{muder}
  \del_\trho \wt{A}_t = \td\ \frac{\tw^4 - \tw_H^4}{\tw^4 + \tw_H^4} \sqrt{\frac{ 1+(\tL')^2 }{\tK}} \ ,
\end{eqnarray}
where $\wt{A}_t := \frac{\sqrt{2}2\pi\a'A_t}{R\sqrt{2B}}$.

There is a trivial solution of (\ref{muder}) with $\td =0$ and
constant $\wt{A}_t$~\cite{Myers3}. The embeddings are then the
same as those at zero chemical potential. For a finite $\td$,
$\tilde{A}_t'$ is singular at $\trho=0$ and requires a source. In
other words the electric displacement must end on a charge source.
The source is the end point of strings stretching between the D7
brane and the black hole horizon. The string tension pulls the D7
branes to the horizon resulting in black hole
embeddings~\cite{Myers1}.  For such an embedding the chemical
potential($\tmu$) is defined as~\cite{Myers1,Kim}
\begin{eqnarray}
  \tmu &:=&\lim_{\trho \ra \infty} \wt{A}_t(\trho) \nn \\
  &=& \int_{\trho_H}^\infty d\trho\ \td\ \frac{\tw^4 - \tw_H^4}
  {\tw^4 + \tw_H^4} \sqrt{\frac{ 1+(\tL')^2 }{\tK}} \ , \label{mu}
\end{eqnarray}
where we fixed $\wt{A}_t(\trho_H) = 0$ for a well defined $A_t$
at the black hole horizon.

The Euclideanized on shell bulk action can be interpreted as the
thermodynamic potential of the boundary field theory. The Grand
potential ($\tOmega$) is associated with the action
(\ref{OriginalAction}) while the Helmholtz free energy ($\tF$) is
associated with the Legendre transformed action
(\ref{LegendreAction}):
\begin{eqnarray}
  && \tF(\tw_H,\td) :=   \frac{-\wt{S}}{N_f T_{D7} (R\sqrt{B})^4
  \mathrm{Vol}} \nn \\
  && \quad = \int_{\trho_H}^\infty d\trho\  \frac{\tw^4 - \tw_H^4}{\tw^4 }
  \sqrt{\tK(1+(\tL')^2)} \label{F}
\end{eqnarray}
\begin{eqnarray}
  && \tOmega(\tw_H,\tmu) :=  \frac{-{S}}{N_f T_{D7} (R\sqrt{B})^4
   \mathrm{Vol}} \nn \\
  && \quad = \int_{\trho_H}^\infty d\trho\  \frac{\tw^4 - \tw_H^4}{\tw^4 }
  \sqrt{\frac{(1+(\tL')^2)}{\tK}} \times \nn \\
  && \qquad \qquad \qquad \left(\left(\frac{\tw^4+\tw_H^4}{\tw^4}\right)^2
   \trho^6 +\frac{1}{\tw^4} \trho^6\right) \label{Omega}
\end{eqnarray}
where $\mathrm{Vol}$ denote the trivial 7-dimensional volume integral
except $\trho$ space, so the thermodynamic potentials defined
above are densities, strictly speaking. Since $\tK \sim \trho^6$,
both integrals diverge as $\trho^3$ at infinity and need to be
renormalized. Thermodynamic potentials, (\ref{mu}),(\ref{F})
and (\ref{Omega}) are reduced to $B=0$ case
if we simplify omit all $\frac{\trho^6}{\tw^4}$ and then tildes.
See for example (\ref{Omega1}).

\section{Chiral Symmetry Breaking and the Thermal Phase Transition}

We begin by reviewing the results of \cite{Johnson1,Erdmenger1} on
magnetic field induced chiral symmetry breaking and the thermal
phase transition to a phase in which the condensate vanishes.
While they show the embeddings for fixed $T$ and different values
of $B$, we will show the embeddings for fixed $B$ and different
values of $T$. By fixing $B$ we are using it as the intrinsic
scale of symmetry breaking in the same fashion as $\Lambda_{QCD}$
plays that role in QCD.

\begin{figure*}[]
\centering
  \subfigure[Low temperature - $\ \tw_H = 0.15$. Here we see chiral symmetry breaking with
  the blue embedding thermodynamically preferred over the red at $\tm=0$.]
  {\includegraphics[width=5cm]{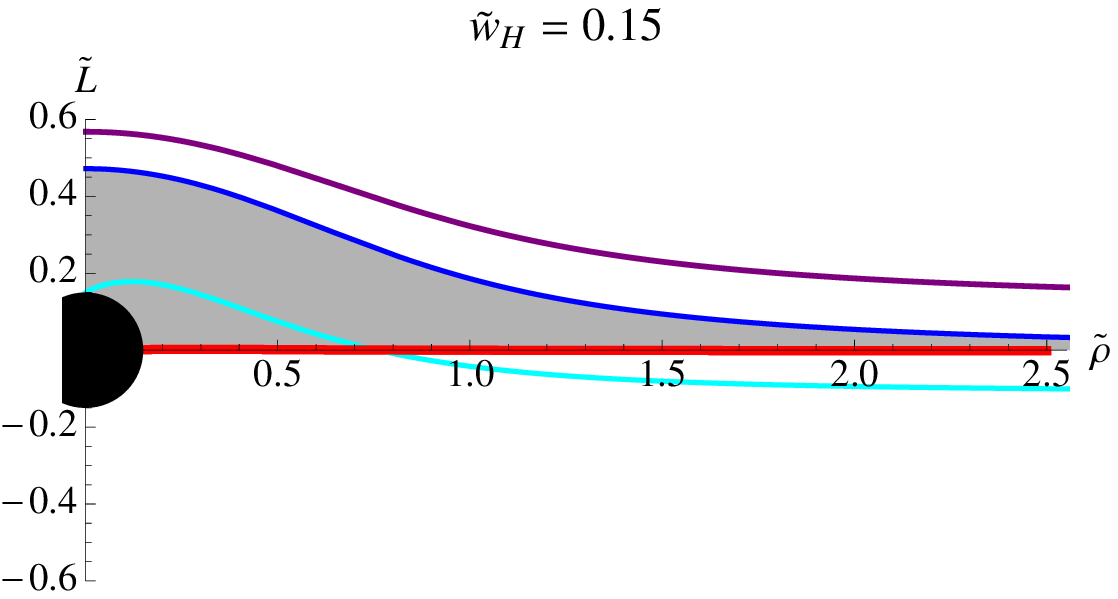}\qquad
   \includegraphics[width=5cm]{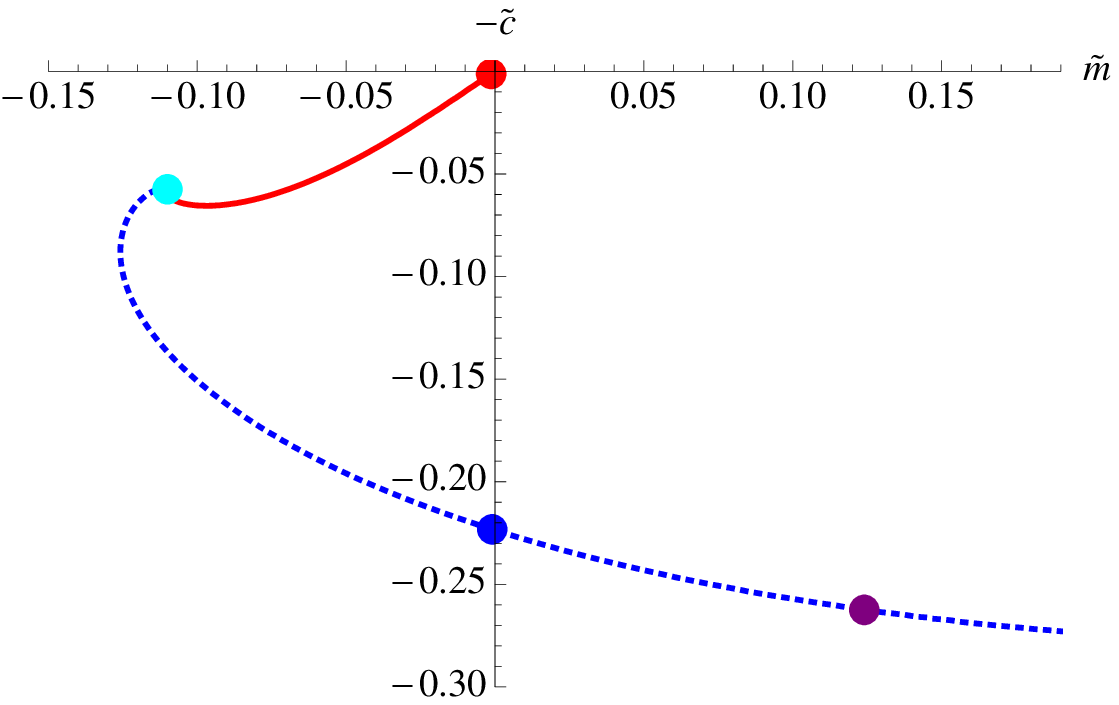}\qquad\quad
   \includegraphics[width=5cm]{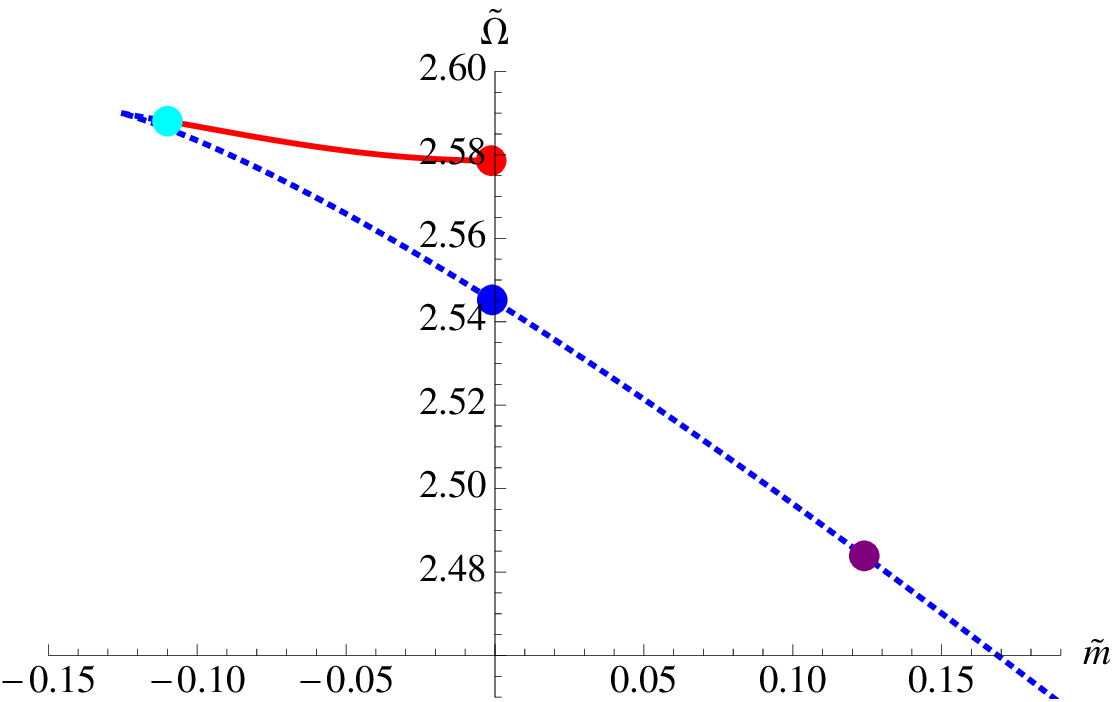}}
  \subfigure[ Transition temperature - $\tw_H = 0.2516$.  This shows the point where
            the first order chiral symmetry phase transition
            occurs from the blue to the red embedding.
            The transition can be identified by considering Maxwell's construction(Middle) or
            the lowest free energy(Right).]
  {\includegraphics[width=5cm]{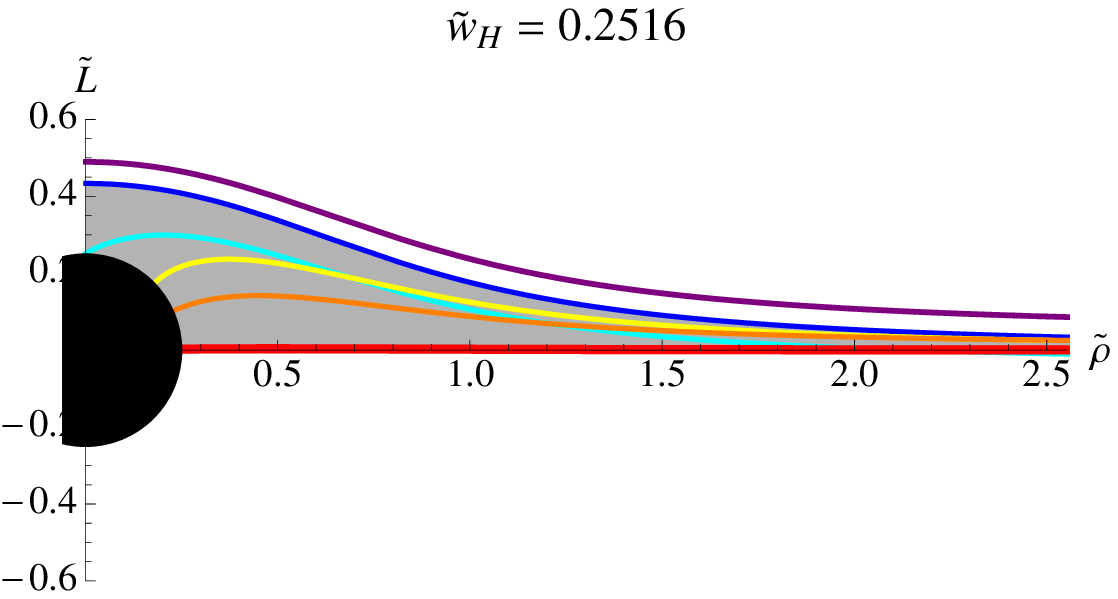}\qquad
   \includegraphics[width=5cm]{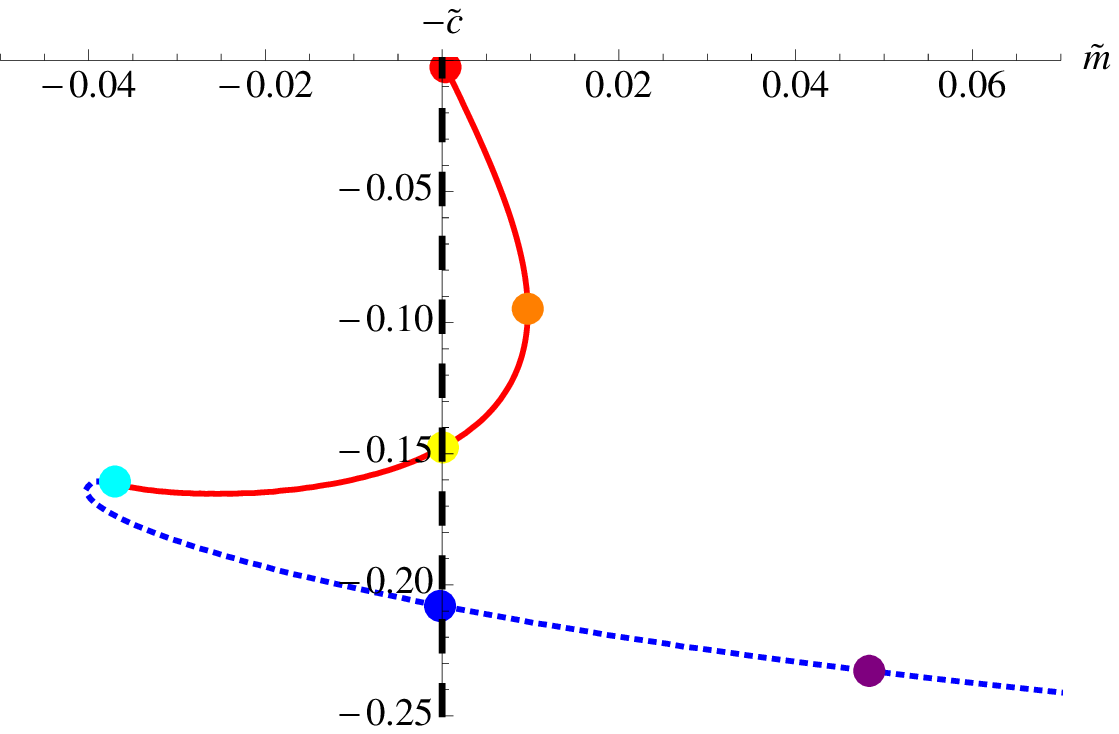}\qquad\quad
   \includegraphics[width=5cm]{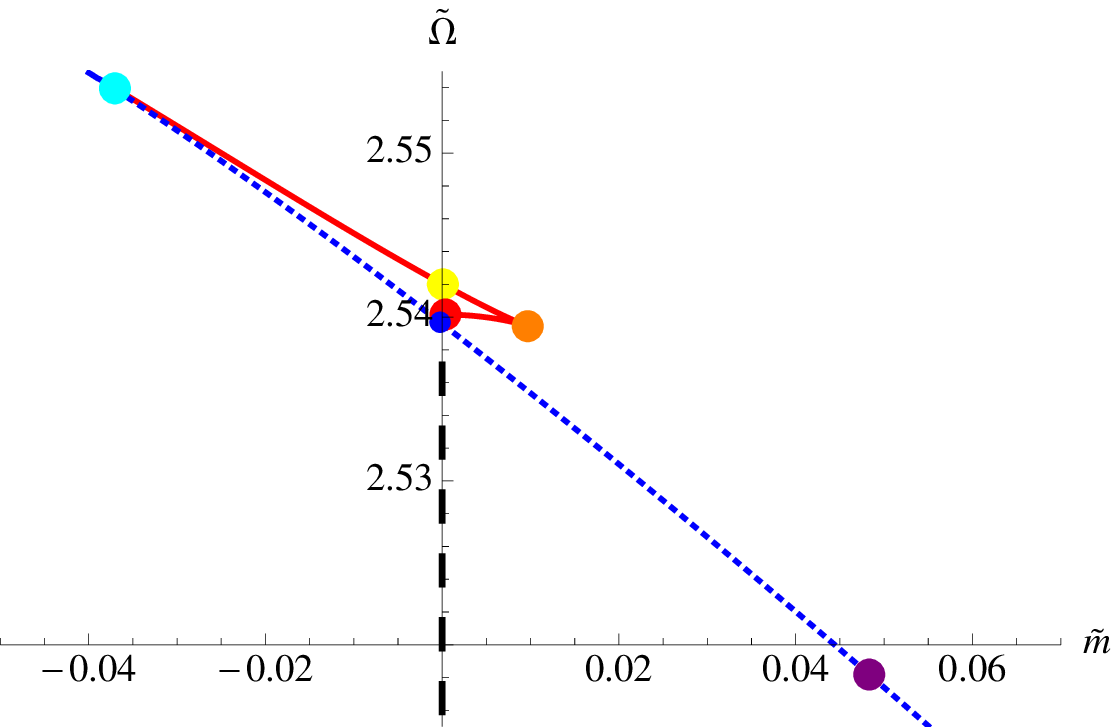}}
  \subfigure[Above the transition - $\ \tw_H = 0.3$. This is the chiral restored phase with the
          $\tm=0$ curve lying along the $\trho$ axis (red).]
  {\includegraphics[width=5cm]{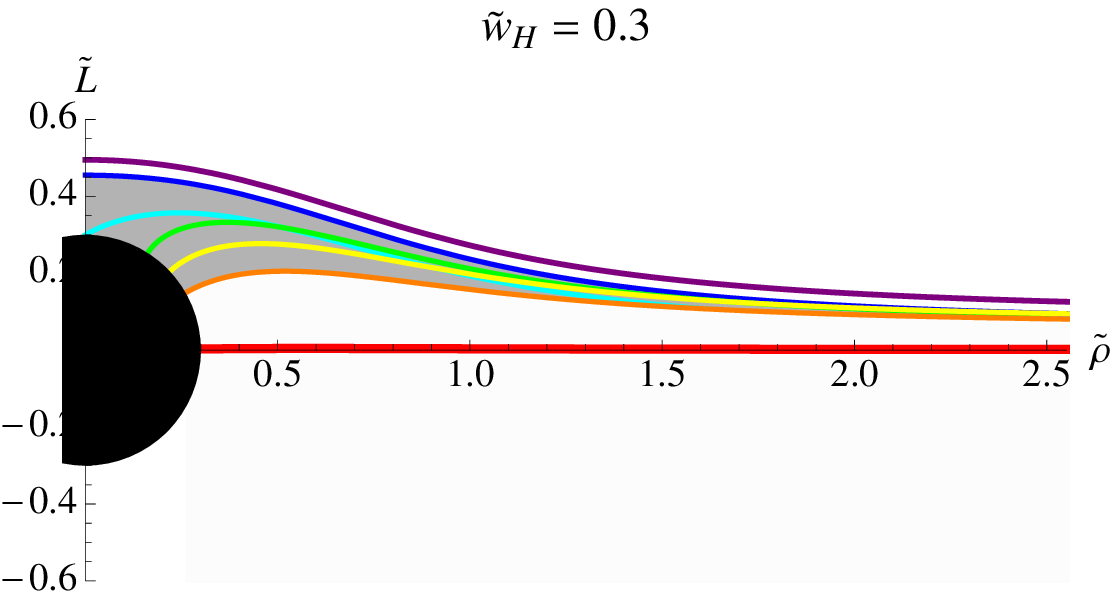}\qquad
   \includegraphics[width=5cm]{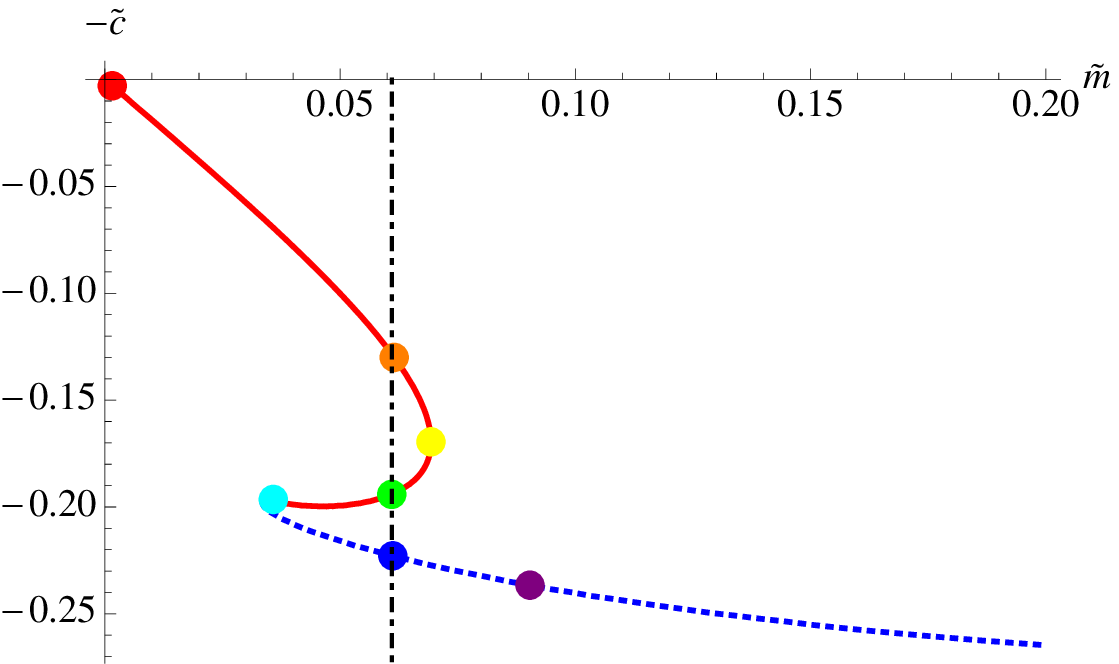}\qquad\quad
   \includegraphics[width=5cm]{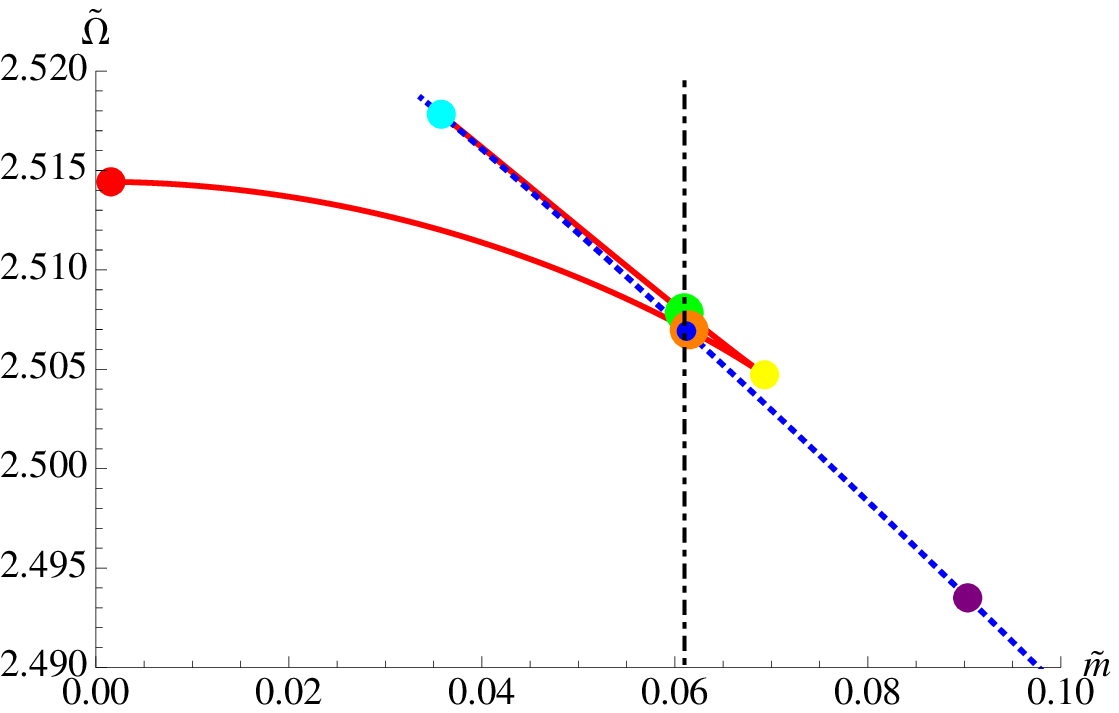}}
  \subfigure[High temperature - $\ \tw_H = 300$.  Here the magnetic field is negligible
           and the embeddings show the usual finite temperature meson
           melting transition.]
  {\includegraphics[width=5cm]{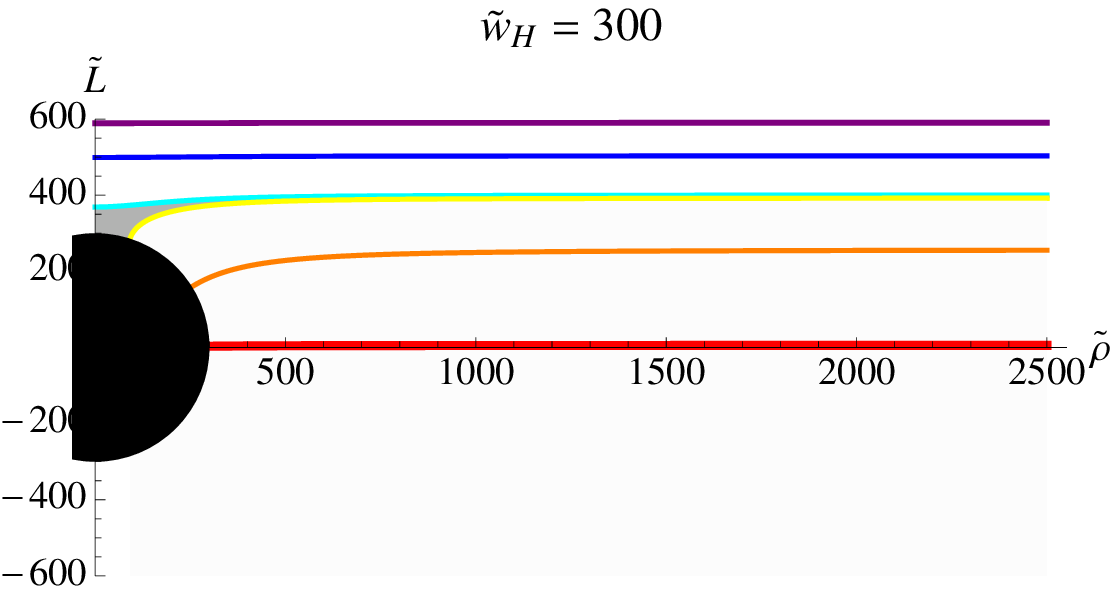}\qquad
   \includegraphics[width=5cm]{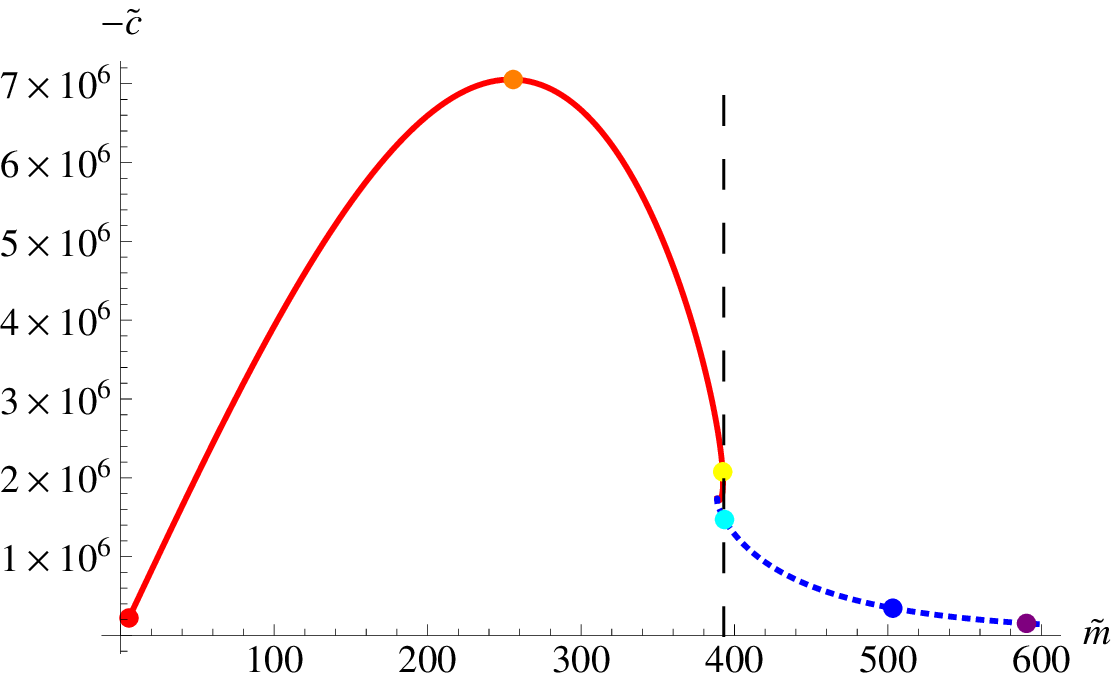}\qquad\quad
   \includegraphics[width=5cm]{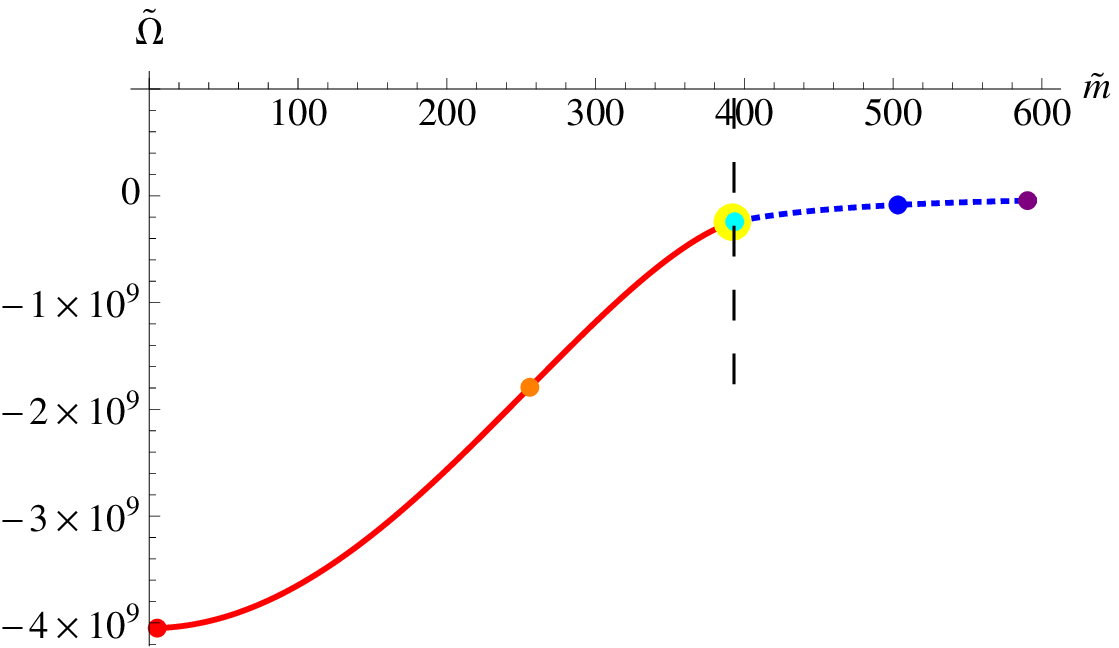}}
  \caption{ % {\Large $\!\!\!\!\!\!\!\!\;
            % \bold{\color{white}{\blacksquare}}\!\!\!\!\!\!\!\! $ }
            % 1a:
           {\small
            The D7 brane embeddings(Left), their corresponding
           $ \tm - \tc $ diagrams(Middle), and the Free energies(Right)
           in the presence of a magnetic field at finite temperature.
           (Parameters are scaled or $B=1/2R^2$
           in terms of parameters without tilde.)}
           }\label{Fig1}
\end{figure*}

Let us digress here to explain how to understand the figures we
will present in this paper. For example, in Fig \ref{Fig1} we have three
columns. The left is the D7 brane embedding configuration. The
middle shows a plot of the allowed values of the condensate $\tc$
as a function of the quark mass $\tm$ - these are thermodynamical
conjugate variables. The right is the corresponding thermodynamic
potential. Each row is for a fixed parameter we are varying - here
it's temperature.
 The left and middle plots are plotted by solving the
equation of motion (\ref{tH}) with the black hole boundary
condition that the embedding is orthogonal to the horizon.

The right hand plot is calculated using (\ref{F}) or
(\ref{Omega}). Both are the same at zero density. We subtract
$\lim_{\trho\ra\infty}\frac{1}{4}\trho^4$ to remove the common
infinite component.

 Every {\it point} in the middle and right plots corresponds to
one embedding {\it curve} in the left plot. These points are color
coded with the colors common across each of the three plots. The
order of colors follows the rainbow from the bottom embedding as a
mnemonic.

 In the middle plot we can find any transition point by a
Maxwell construction (an equal area law), which is also confirmed
by the minimum of the grand potential on the right.  The vertical
dashed line in the middle and right hand plots corresponds to the
transition point.

 In the left plots the gray region contains embeddings that
are excluded since they are unstable, as shown in the middle and
on the right.

The results for the case of a constant magnetic field and varying
temperature are displayed in Fig \ref{Fig1}a-d.
The Fig \ref{Fig1}a (Left) shows the D7 embeddings when $T\ll B$ and the black
hole is small. The embeddings are driven away from the origin of the
$\tL-\trho$ plane - this behaviour is a result of the inverse
powers of $\tw$, when $\tw_H \ll 1 $, in the Lagrangian
(\ref{rescale}) which lead the action to grow if the D7 approaches
the origin (note that the factor of $\trho^3$ multiplying the
action means the action will never actually diverge).
There is also a embedding that end on the black hole (shown
in red) but they are thermodynamically disfavoured as shown in
Fig \ref{Fig1}a (Right).

At large $\trho$ the stable embedding with $\tm=0$ has a non-zero
derivative so $\tc$ is non-zero and there is a chiral condensate
\ie $\,$chiral symmetry breaking. The U(1) symmetry in the $\Omega_1$
direction is clearly broken by any particular embedding too.  We
can numerically read off the values of $\tm$ and $\tc$ from the
embeddings and their values are shown
in Fig \ref{Fig1}a (Middle), where the dotted blue curves
are for Minkowski embeddings, whilst the red curves are for
black hole embeddings.

If the temperature is allowed to rise sufficiently then the black
hole horizon grows to mask the area of the plane in which the
inverse $\tw$ terms in the Lagrangian are large. At a critical
value of $T$ the benefit to the $\tm=0$ embedding of curving off
the axis becomes disfavoured and it instead lies along the
$\tilde{\rho}$ axis - chiral symmetry breaking switches off. This
first order transition occurs at $\tilde{w}_H = 0.2516$ as shown in
Fig \ref{Fig1}b by Maxwell's construction (Middle) and by lower
grand potential (Right). Our value for the
critical temeprature agrees with the value $\wt{B} = 16$ in
\cite{Erdmenger1} since our $\tw$ is the same as
$\sqrt{\frac{1}{\wt{B}}}$ in \cite{Erdmenger1}.

We show an example
of the embeddings above the critical temperature, their grand potential
and the evolution of the curves in the $\tm-\tc$ plane in Fig \ref{Fig1}c.

The Fig \ref{Fig1}d shows a case when $T \gg B$ when the area of
the plane in which B is important is totally masked by the black
hole and the results match those of the usual finite T version of
the ${\cal N}$=2 theory. For $\tm > \tilde{u}_H$ the embeddings
are Minkowski like whilst for small $\tm$ they fall into the black
hole. There is a first order phase transition between these two
phases which is the meson melting phase transition discussed in
detail in
\cite{Babington,Myers2,Apreda:2005yz,Albash:2006ew,Mateos0}.
Minkowski embeddings have a stable mesonic spectrum \cite{Mateos}
whilst in the case of black hole embeddings the black holes'
quasi-normal modes induce an imaginary component to the meson
masses \cite{Peeters:2006iu,Hoyos:2006gb}.  We can see that the
previously reported ``meson melting" transition at large quark
mass becomes also the chiral symmetry restoring transition at zero
quark mass.

\section{Finite density or chemical potential at zero temperature}

\begin{figure*}[]
\centering
  \subfigure[Low density - $\ \td = 0.01$. Here we see chiral symmetry breaking
  (the blue embedding is preferred over the red embedding) and a
  spiral structure in the $\tm$ vs $\tc$ plane.]
  {\includegraphics[width=5cm]{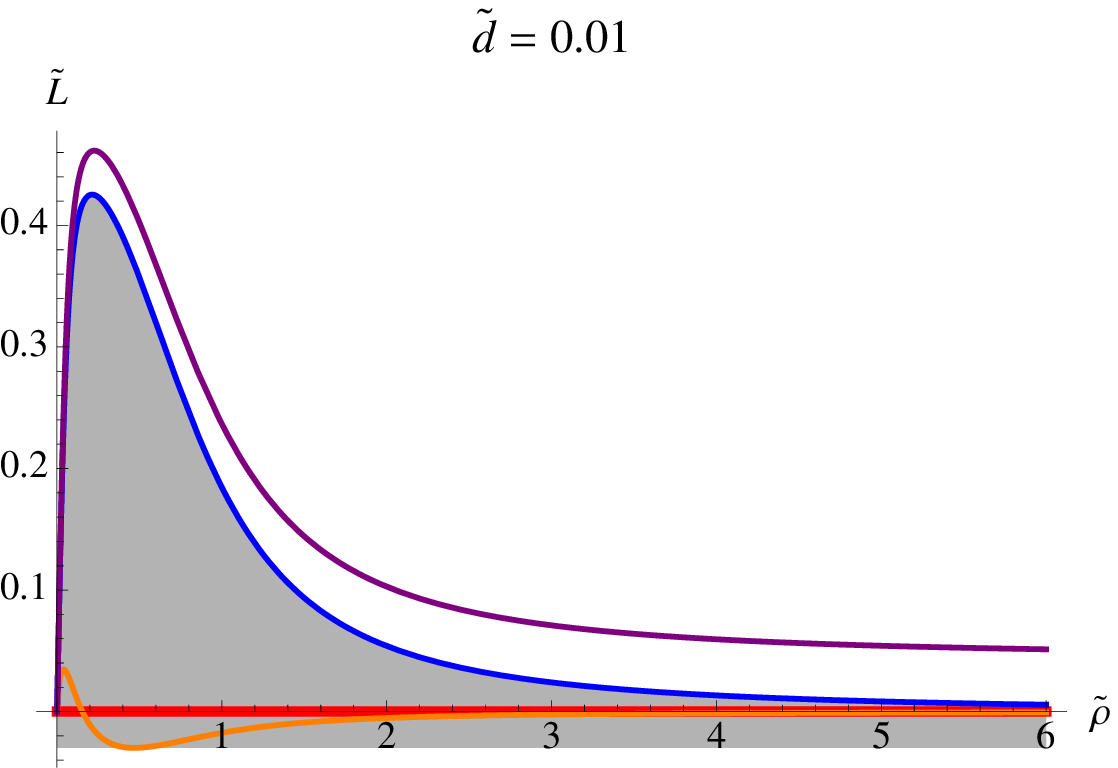}\qquad
   \includegraphics[width=5cm]{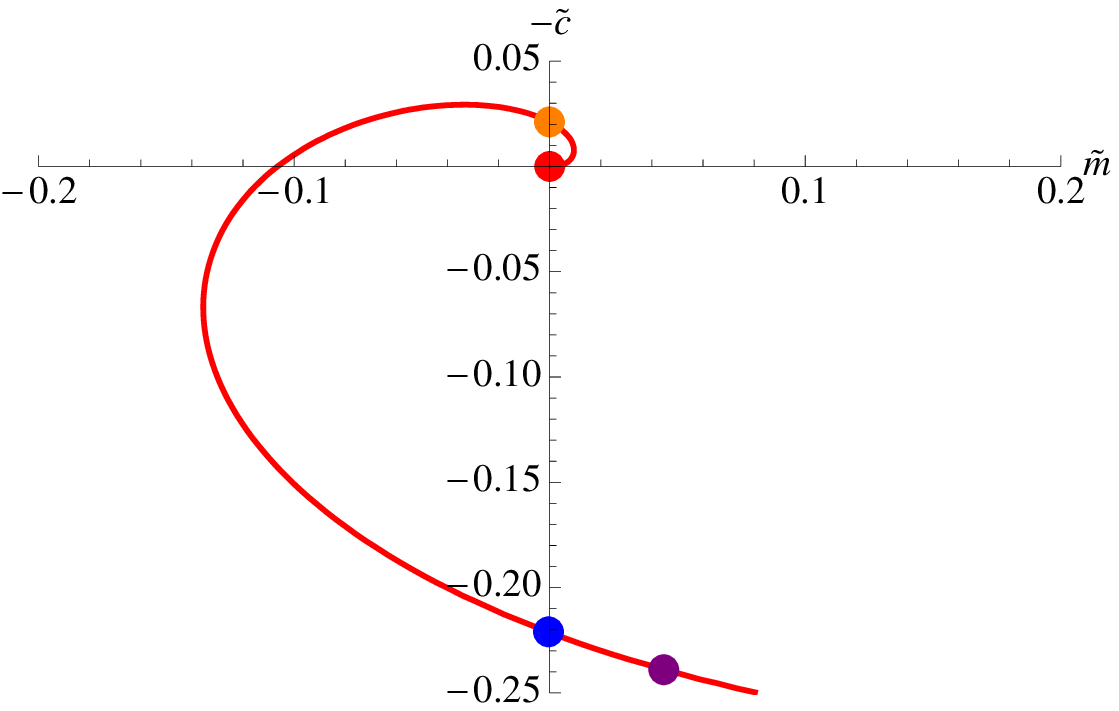}\qquad\quad
   \includegraphics[width=5cm]{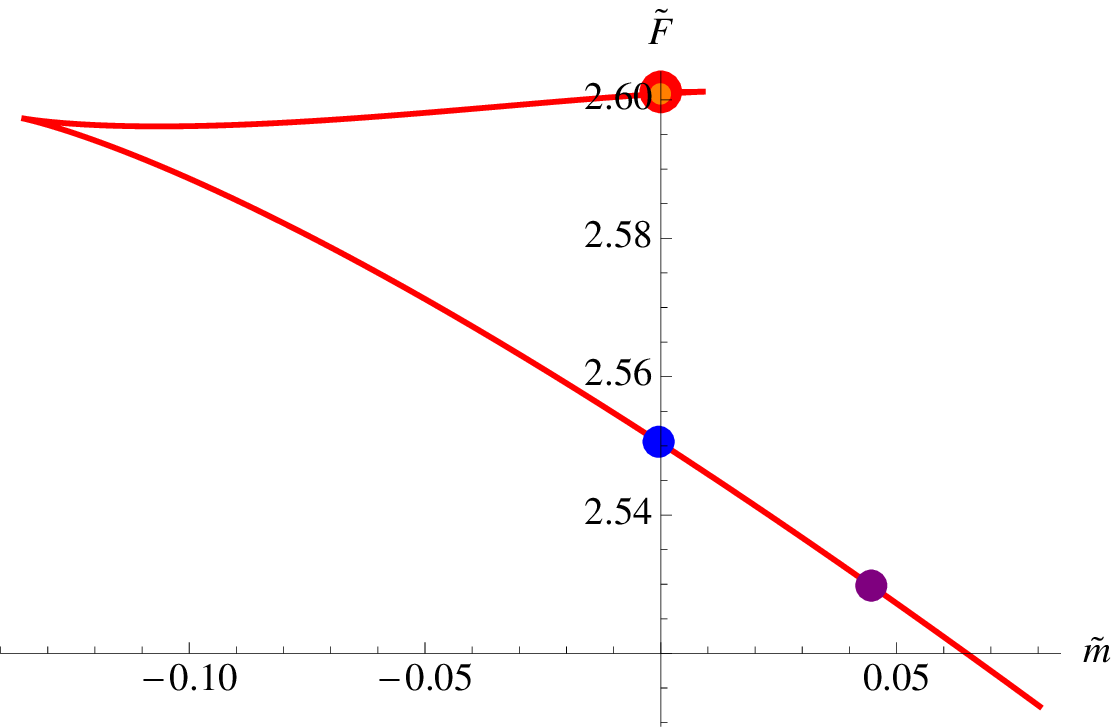}}
  \subfigure[Increasing density below the transition - $\
  \td=0.1$. There is still chiral symmetry breaking here with the
  orange embedding preferred to the red.
  Note the spiral structure in the $\tm- \tc$ plane has
   disappeared.]
  {\includegraphics[width=5cm]{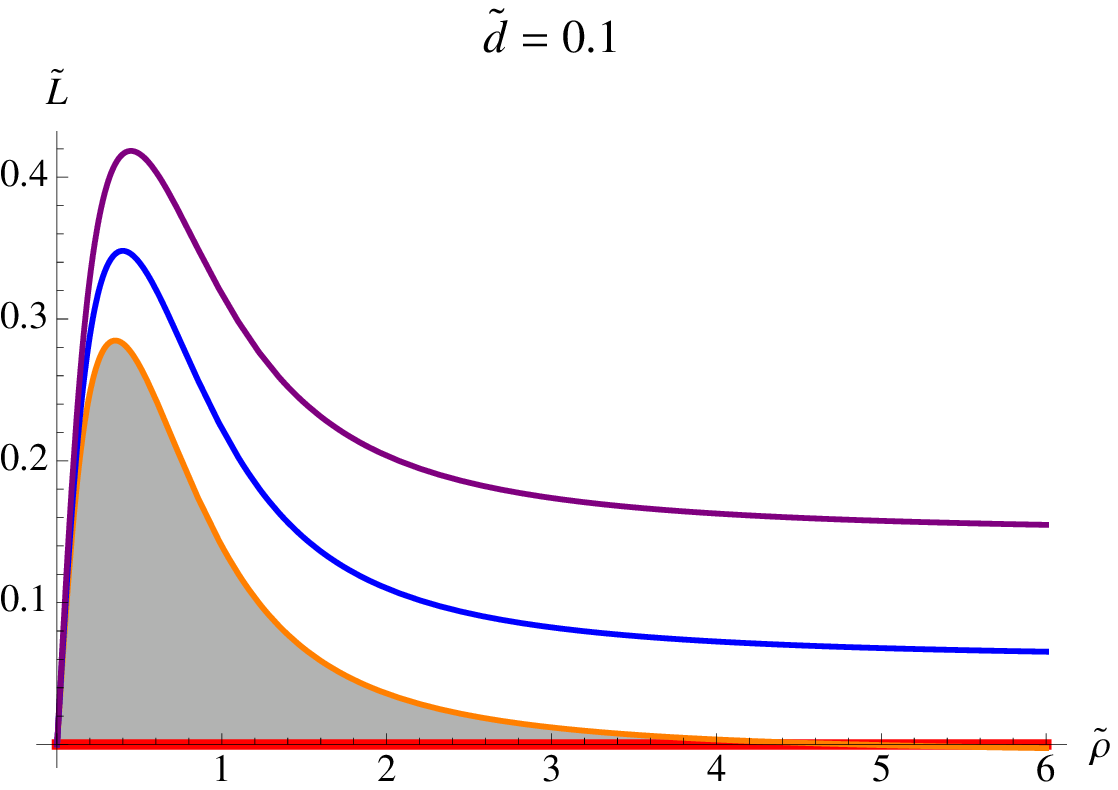}\qquad
   \includegraphics[width=5cm]{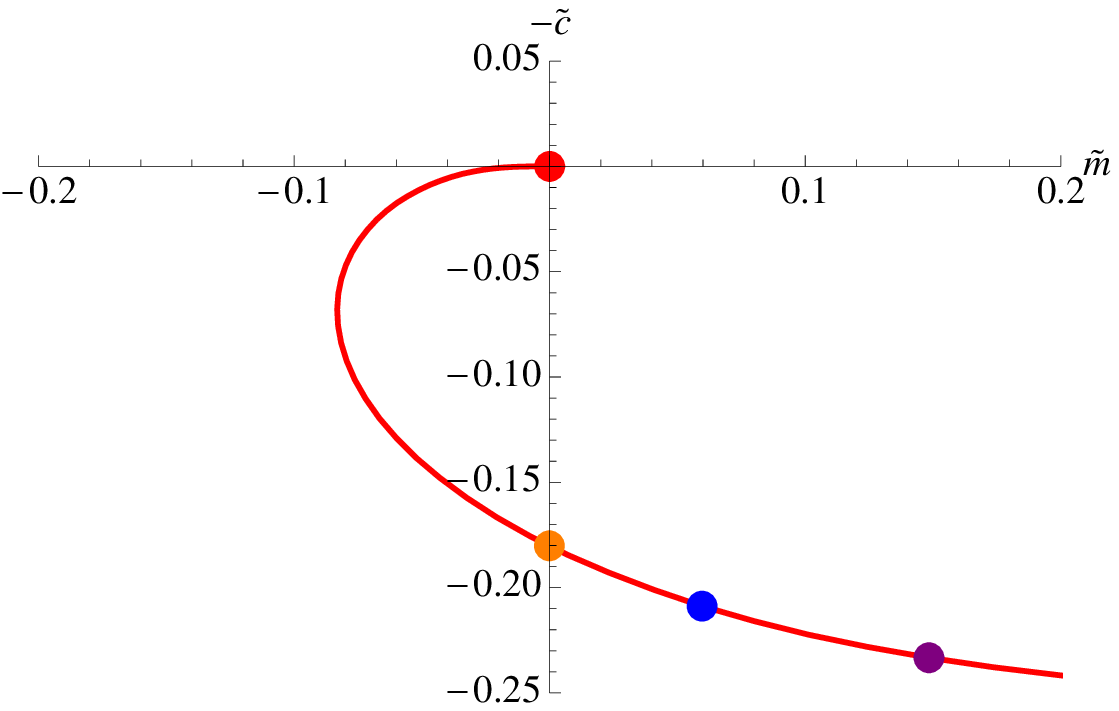}\qquad\quad
   \includegraphics[width=5cm]{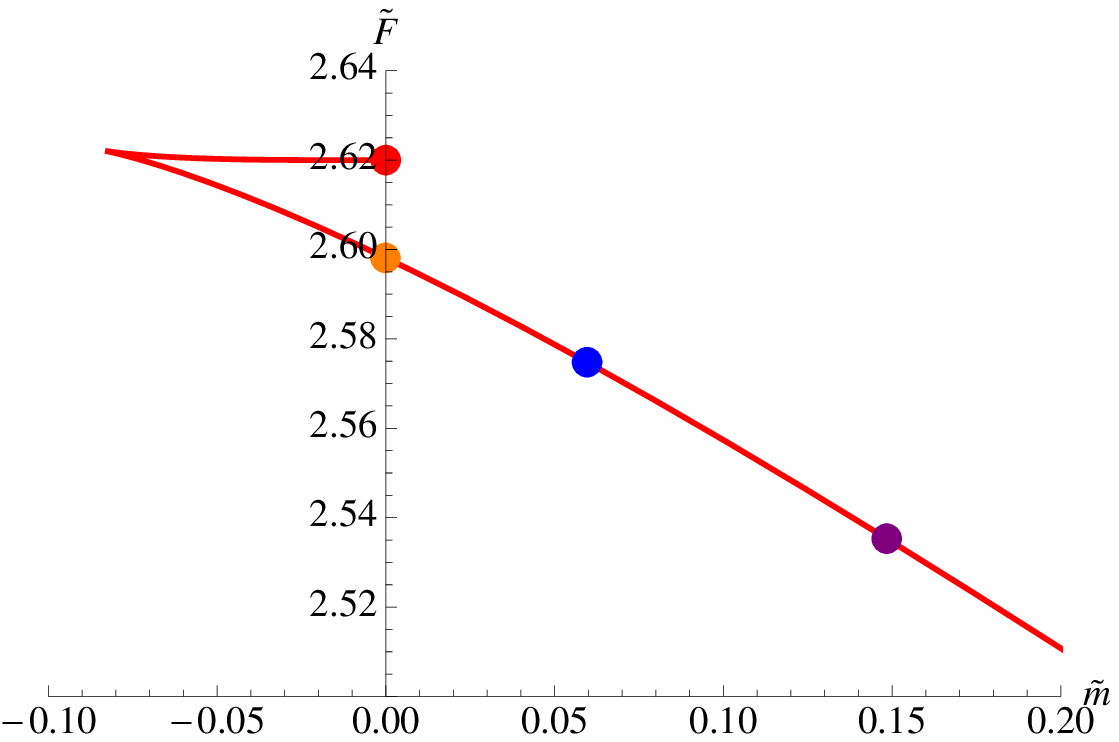}}
  \subfigure[Transition point - $\ \td = 0.3197$.  This shows the point where
            the second order chiral symmetry phase transition occurs.]
  {\includegraphics[width=5cm]{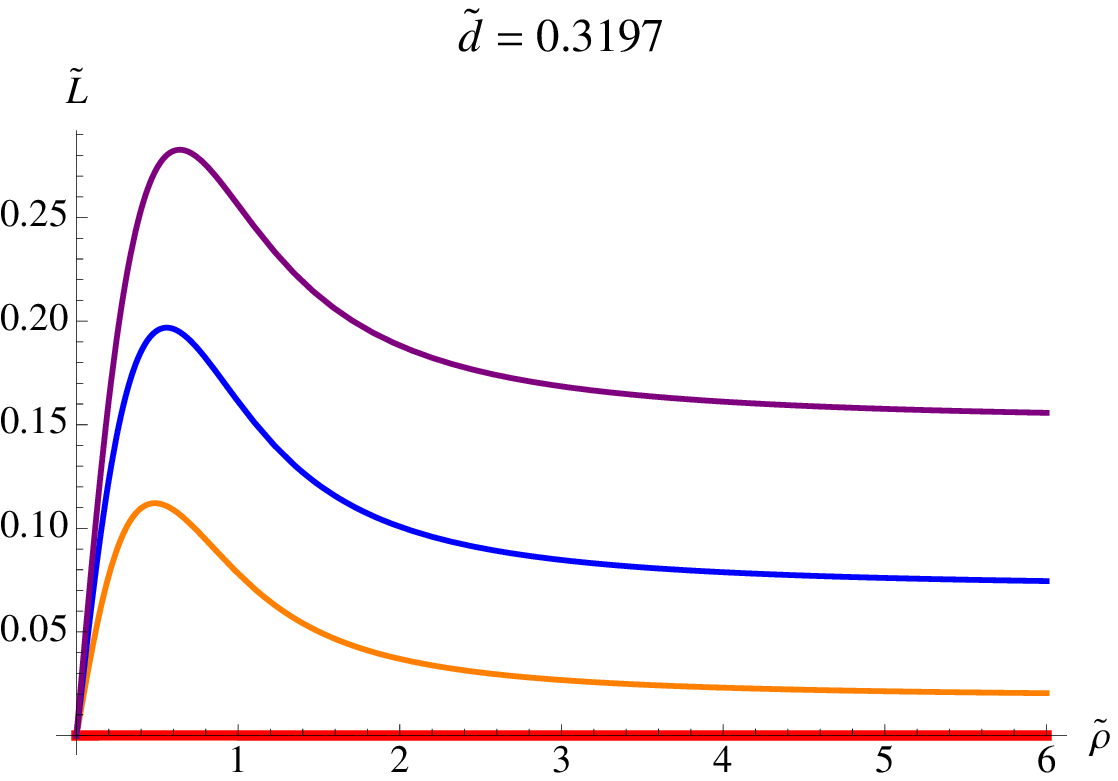}\qquad
   \includegraphics[width=5cm]{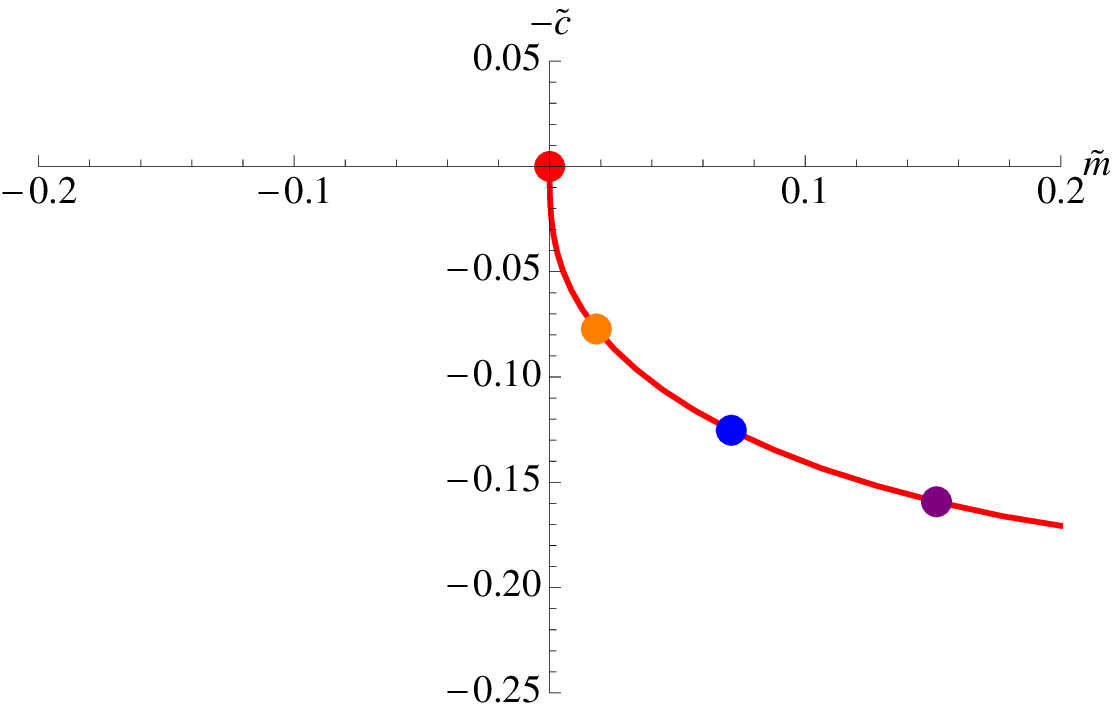}\qquad\quad
   \includegraphics[width=5cm]{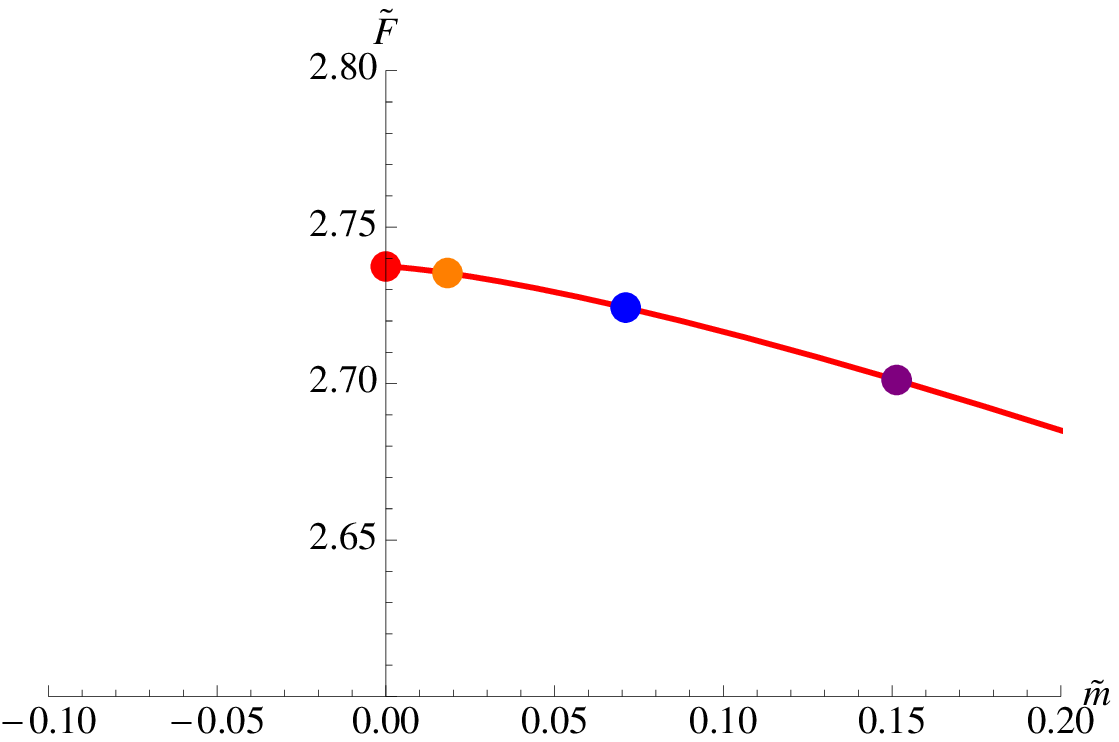}}

  \subfigure[High density $\ \td = 1$. This is the chiral restored phase with the
          $\tm=0$ curve lying along the $\trho$ axis.
          For larger $\tm$ the usual spike like embedding can be seen.]
  {\includegraphics[width=5cm]{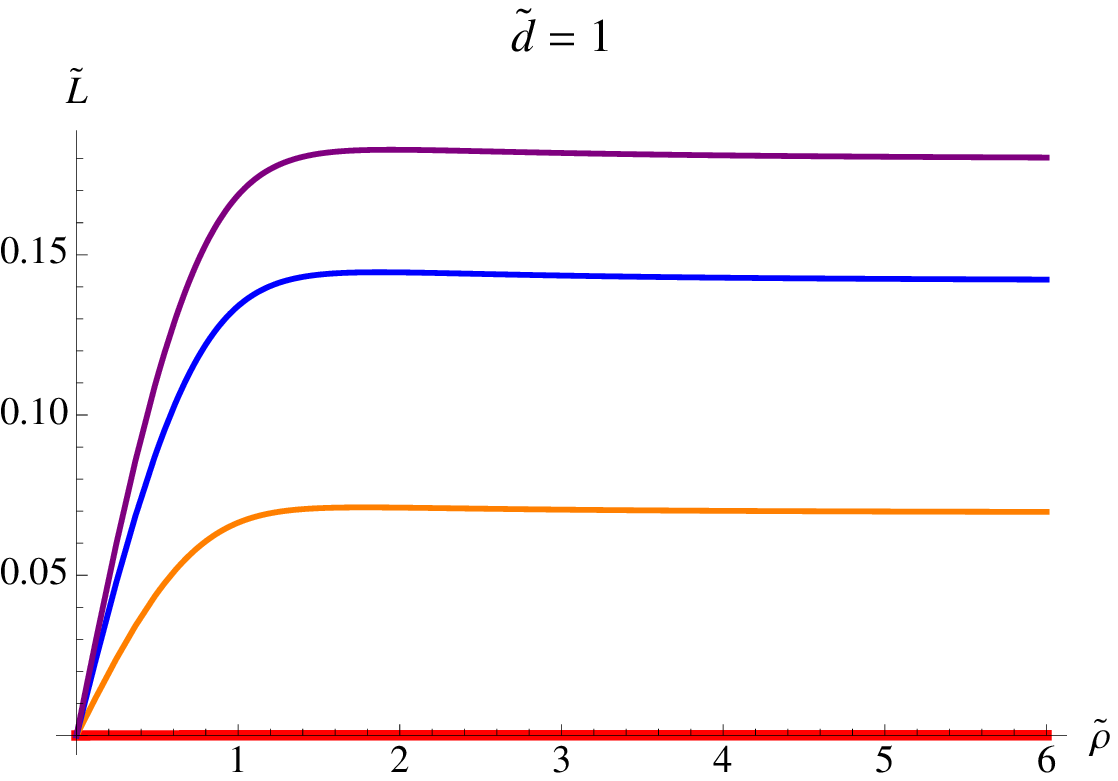}\qquad
   \includegraphics[width=5cm]{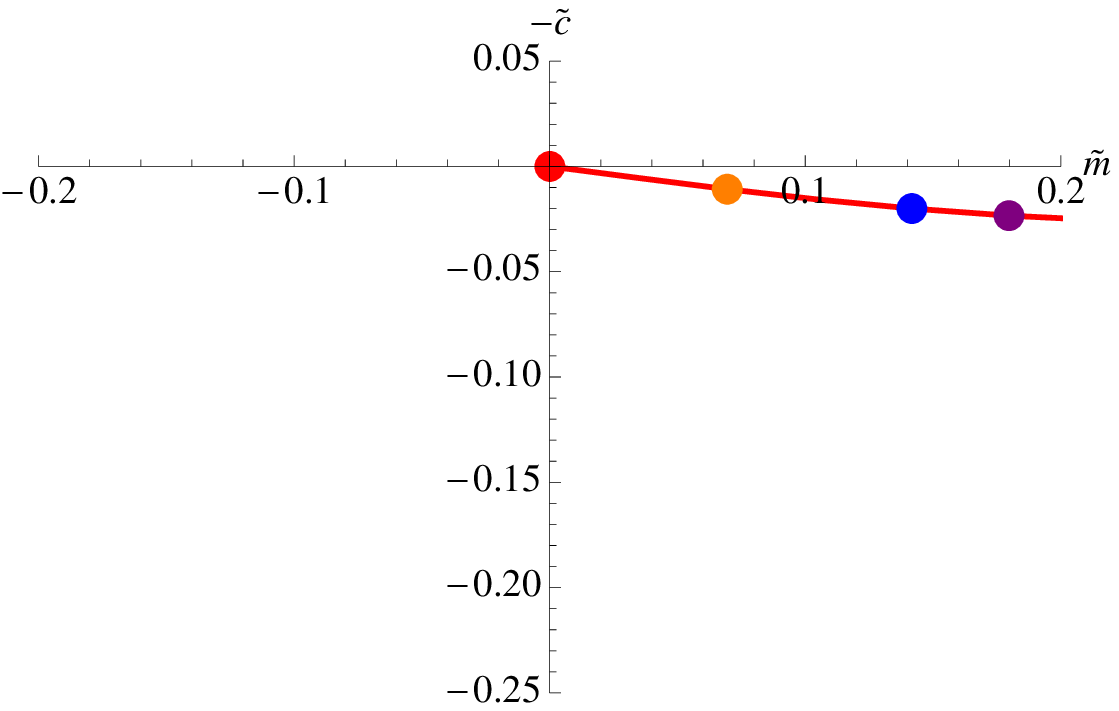}\qquad\quad
   \includegraphics[width=5cm]{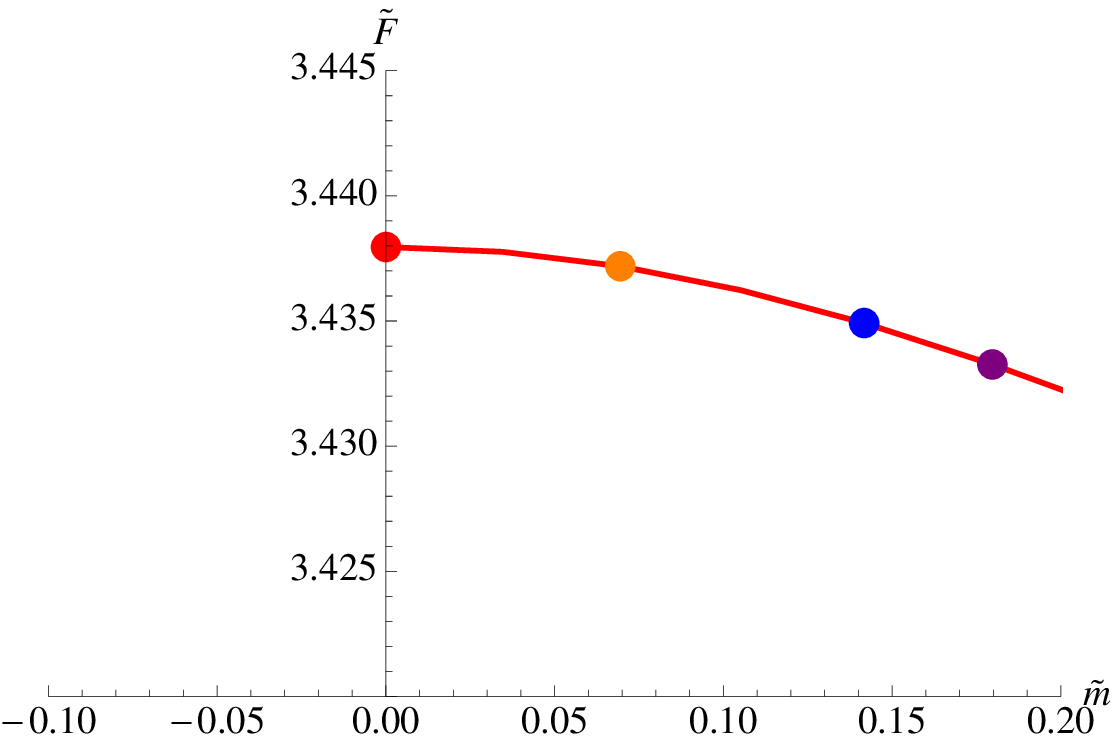}}
  \caption{ % {\Large $\!\!\!\!\!\!\!\!\;
            % \bold{\color{white}{\blacksquare}}\!\!\!\!\!\!\!\! $ }
            % 1a:
           {\small
            The D7 brane embeddings(Left), their corresponding
           $ \tm - \tc $ diagrams(Middle), and the Free energies(Right)
           in the presence of a magnetic field at finite density.
           (Parameters are scaled or $B=1/2R^2$
           in terms of parameters without tilde.)}
           }\label{Fig2}
\end{figure*}
We can now turn to the inclusion of finite density or chemical
potential in the theory with magnetic field.  In this section we
consider the zero temperature ($\tilde{u}_H=0$) theory only, and
will continue to finite temperature in the next section.

A finite density (chemical potential) at zero temperature has been
studied in the ${\cal N}$=2 theory without a magnetic field in
\cite{Karch1}, where analytic solutions for both a black hole like
embedding and a Minkowski embedding have been found. 
When a magnetic field is turned on, analytic solutions are not
available any more, but we have found numerical solutions that
continuously deform from the known analytic solutions at zero
magnetic field.

Minkowski embedding solutions correspond to zero density and 
finite chemical potential.
The black hole like embedding is the embedding deformed by the density - a
spike forms from the D7 down to the origin of the $\wt{L}-\trho$
plane (Fig \ref{Fig2}d (Left))
which has been interpreted as an even distribution of
strings (\ie $\,$ quarks) forming in the vacuum of the gauge
theory.

\begin{figure*}[]
\centering
 \subfigure[Zero temperature - $\ \tw_H = 0$. The second order meson melting
 transition and then the second order chiral restoration transition are apparent. ]
  {\includegraphics[width=5cm]{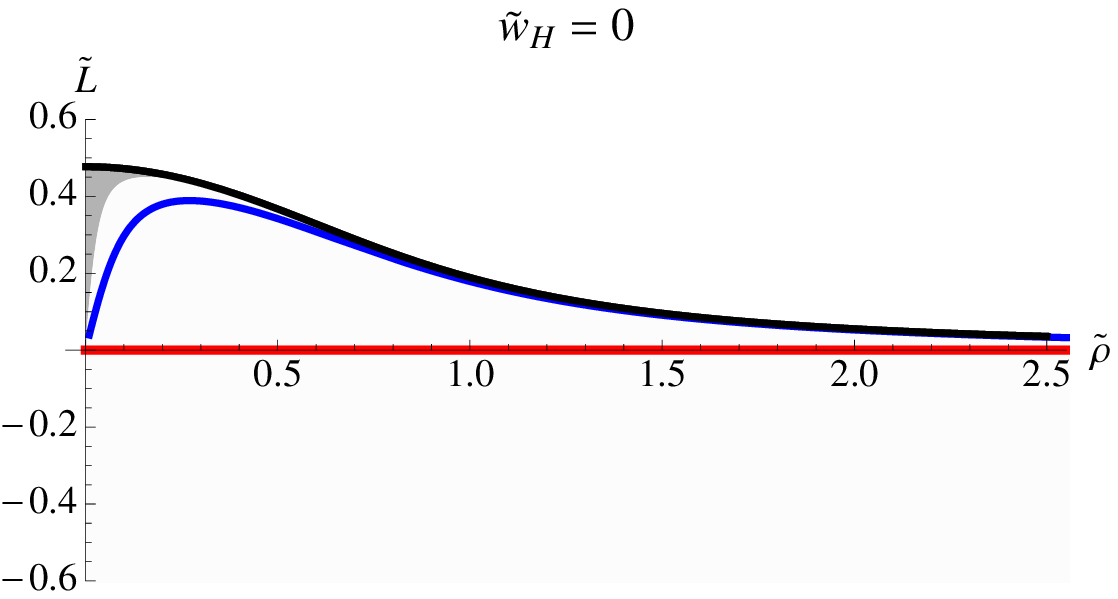}\qquad
   \includegraphics[width=5cm]{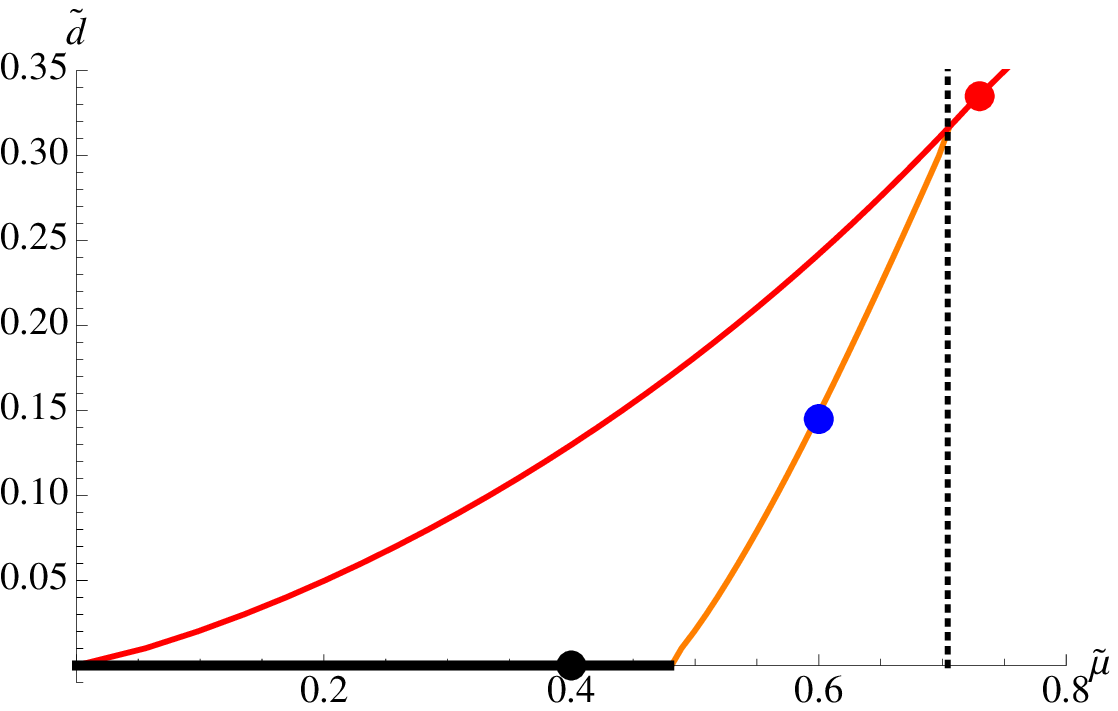}\qquad\quad
   \includegraphics[width=5cm]{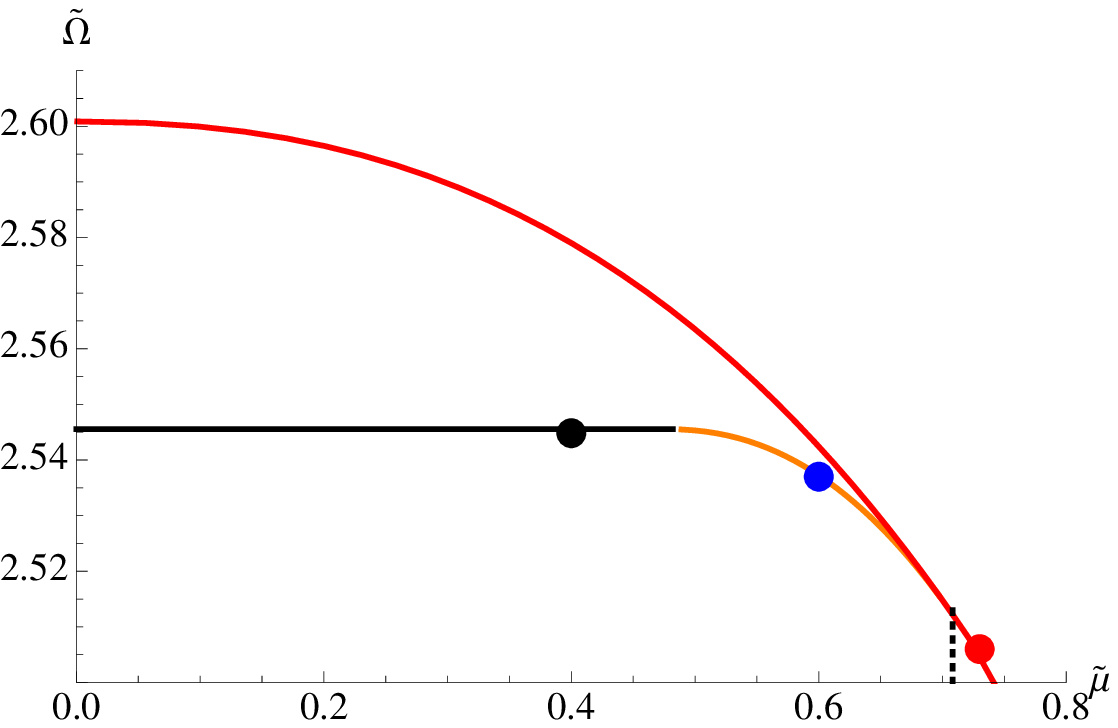}}
  \subfigure[Low temperature - $\ \tw_H = 0.15$. The zero temperature structure remains.]
  {\includegraphics[width=5cm]{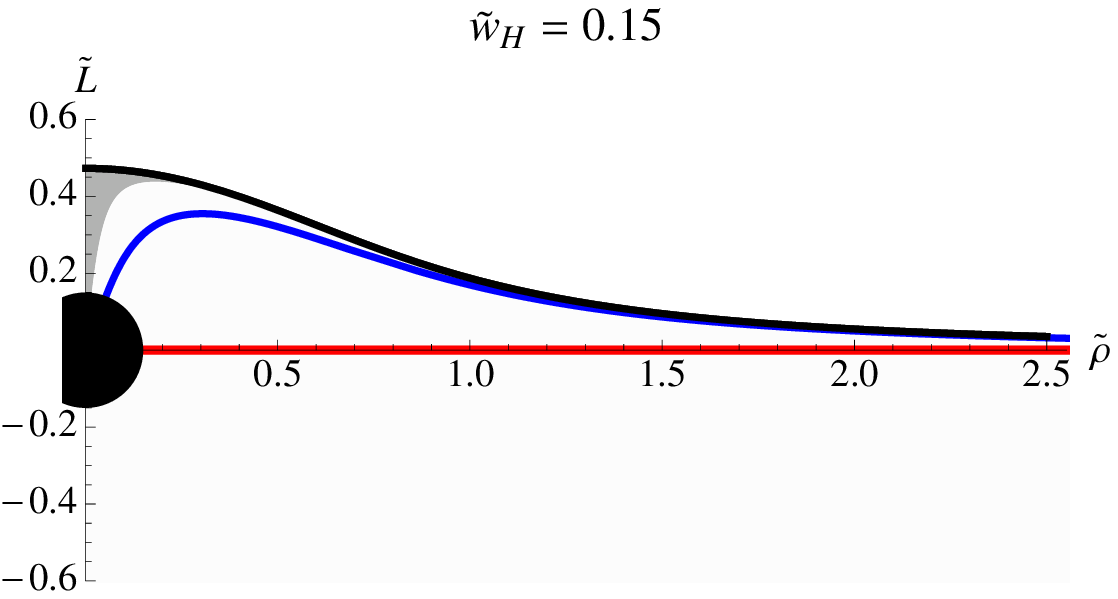}\qquad
   \includegraphics[width=5cm]{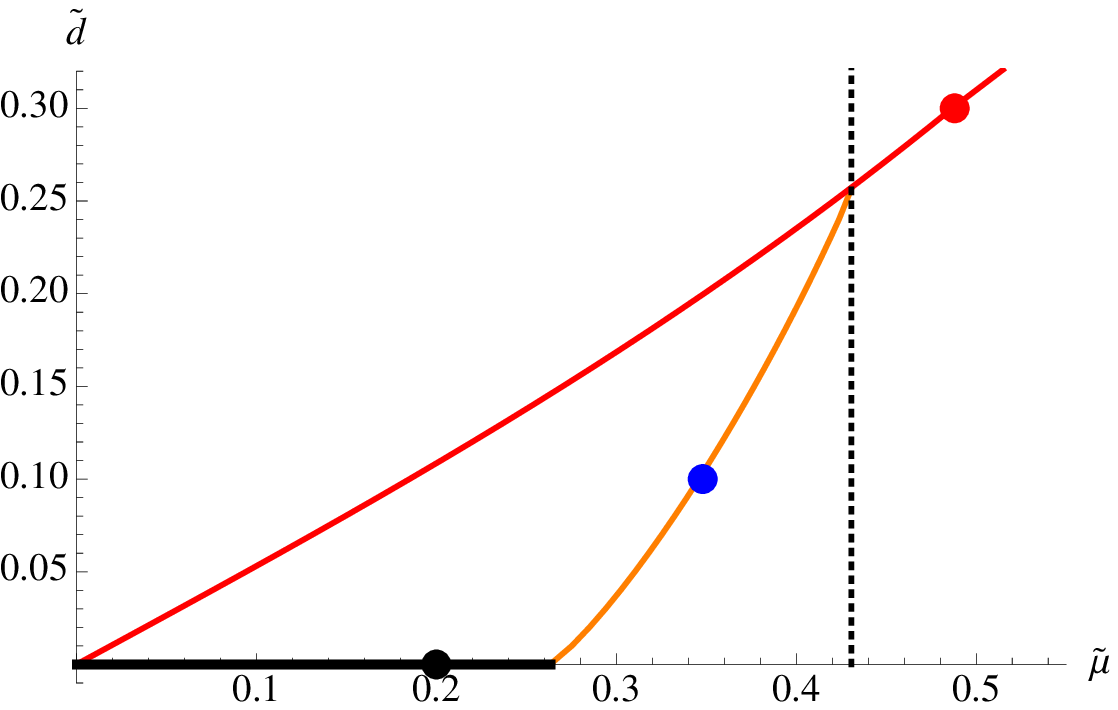}\qquad\quad
   \includegraphics[width=5cm]{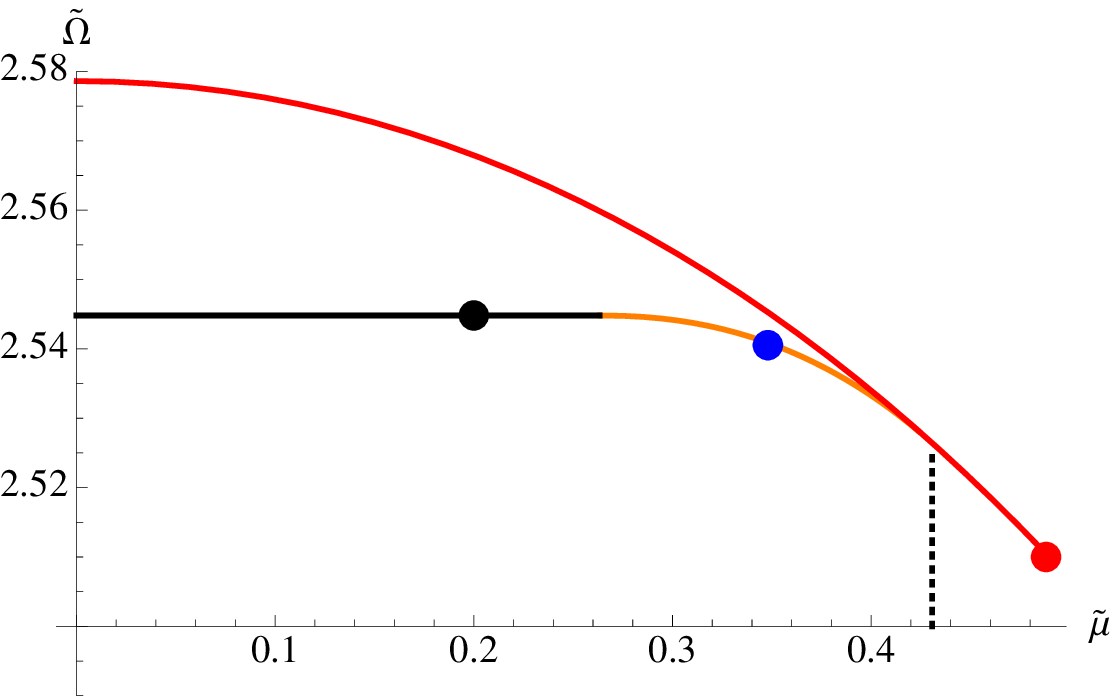}}
  \subfigure[Above the first tri-critical point - $\ \tw_H = 0.23$. The meson melting
   transitions remains second order but the chiral symmetry restoration transition is first order. ]
  {\includegraphics[width=5cm]{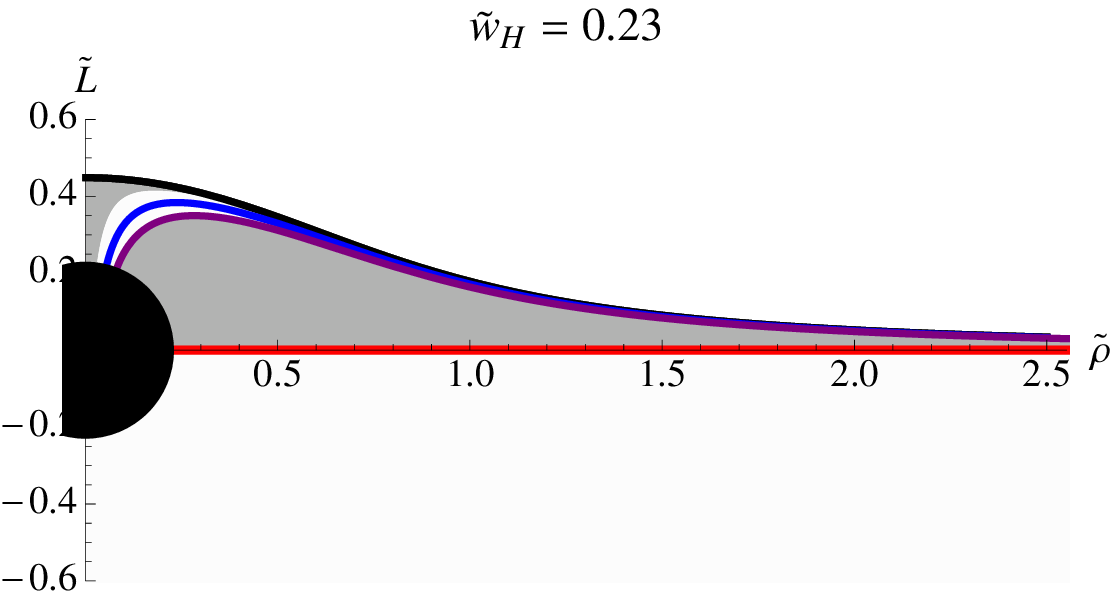}\qquad
   \includegraphics[width=5cm]{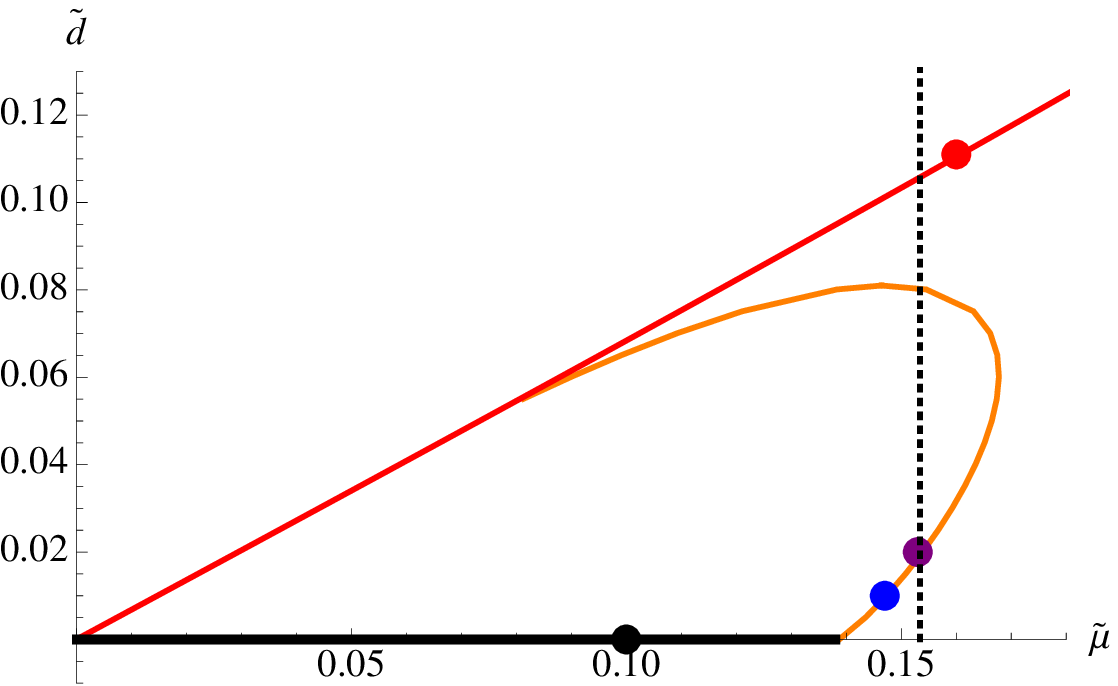}\qquad\quad
   \includegraphics[width=5cm]{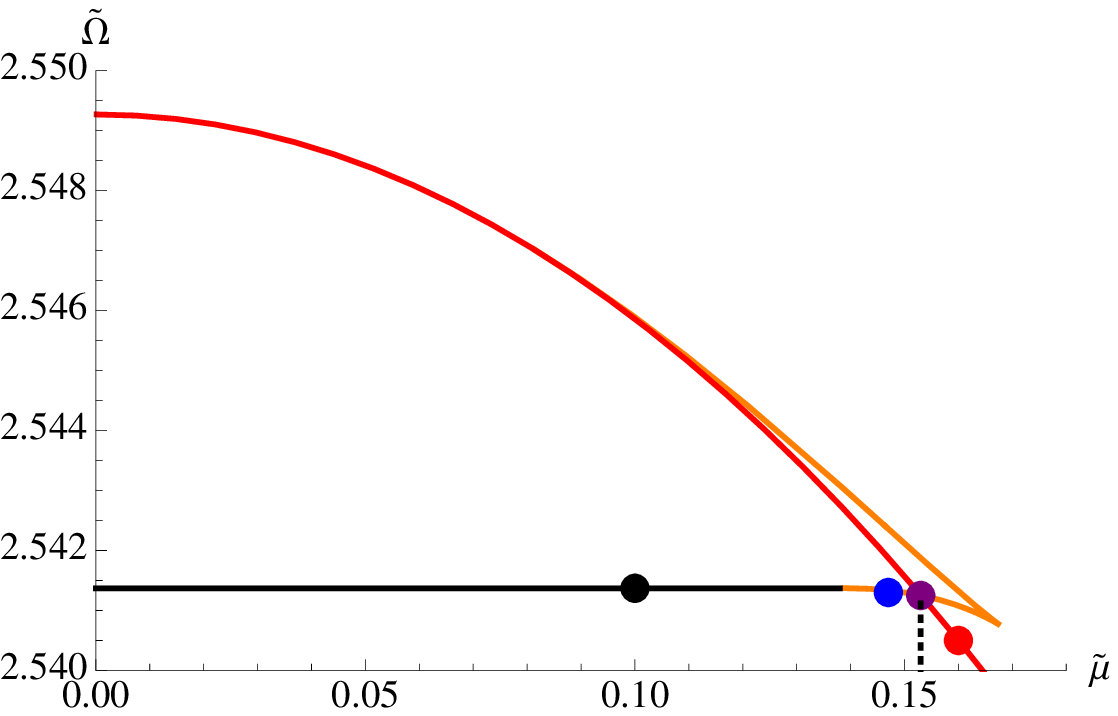}}
  \subfigure[Above the second tri-critical point - $\ \tw_H = 0.24$. There is
  now only a single first order transition for meson melting and chiral symmetry restoration. ]
  {\includegraphics[width=5cm]{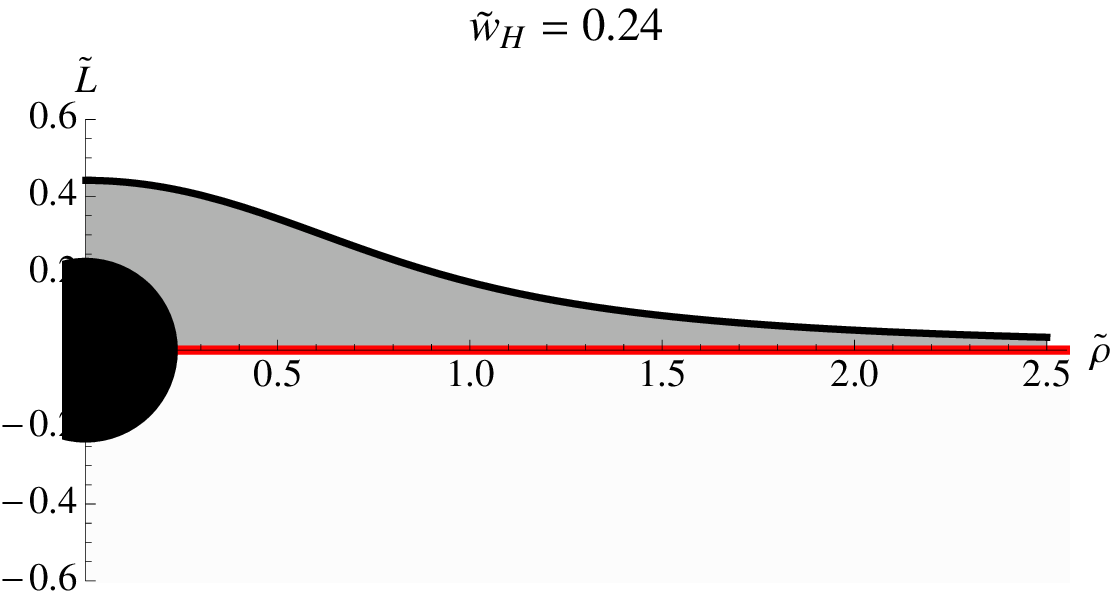}\qquad
   \includegraphics[width=5cm]{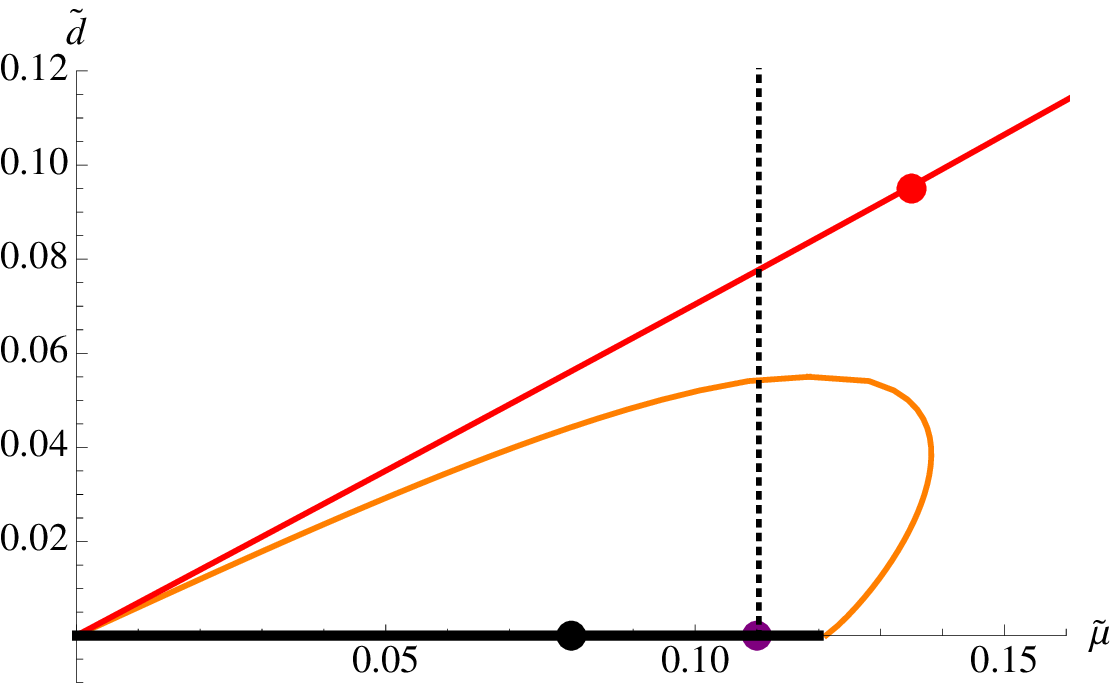}\qquad\quad
   \includegraphics[width=5cm]{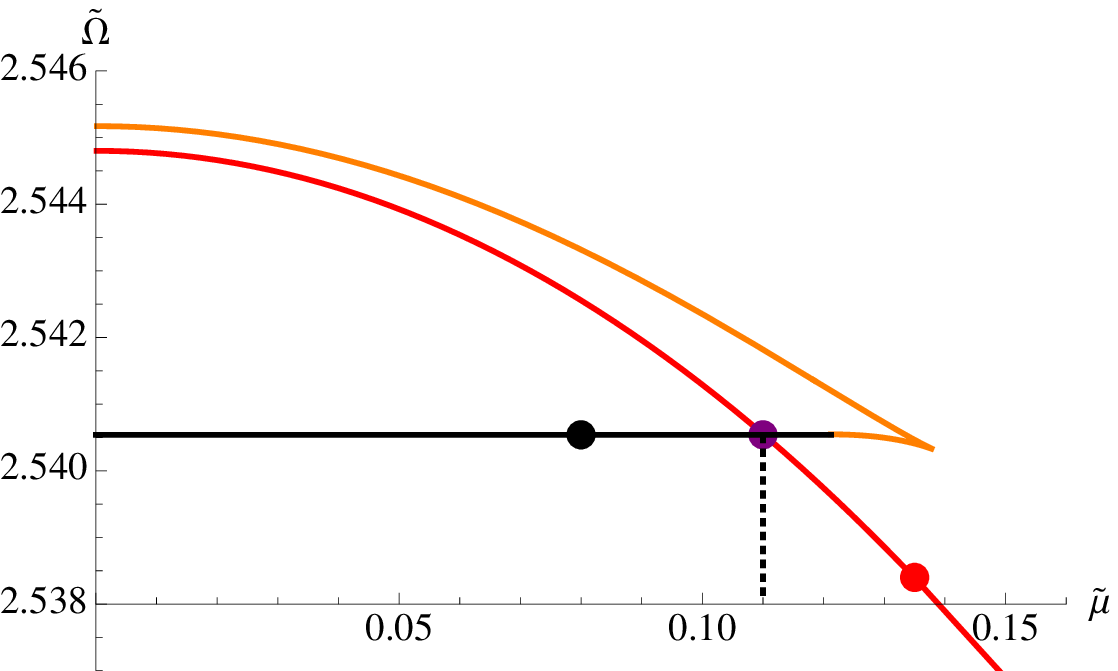}}
  \subfigure[High temperature - $\ \tw_H = 0.2516$. The ground state preserves chiral
  symmetry for all values of $\tilde{\mu}$. ]
  {\includegraphics[width=5cm]{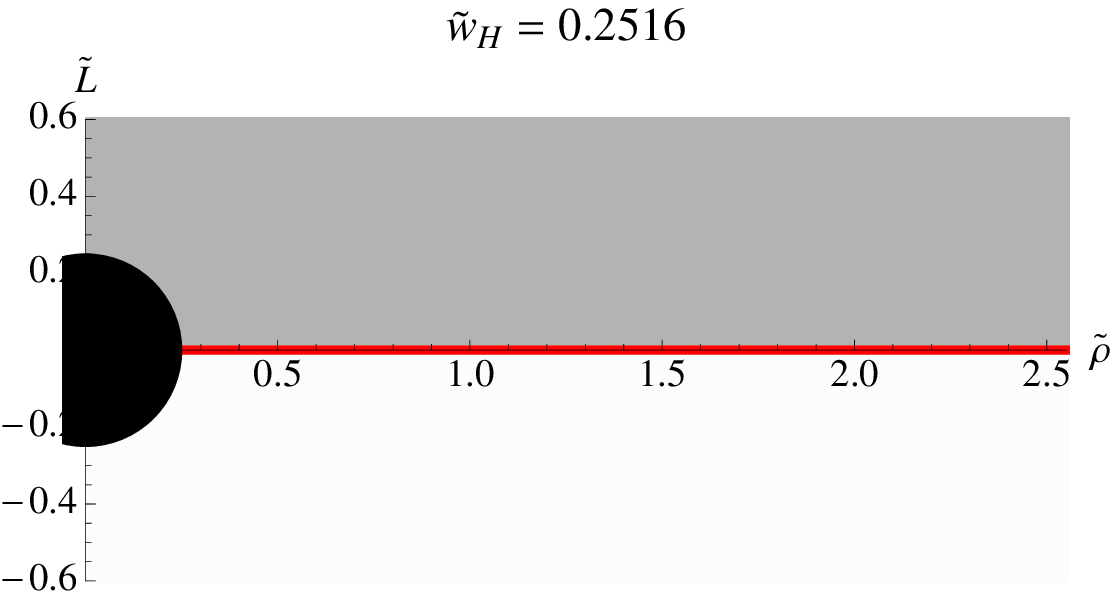}\qquad
   \includegraphics[width=5cm]{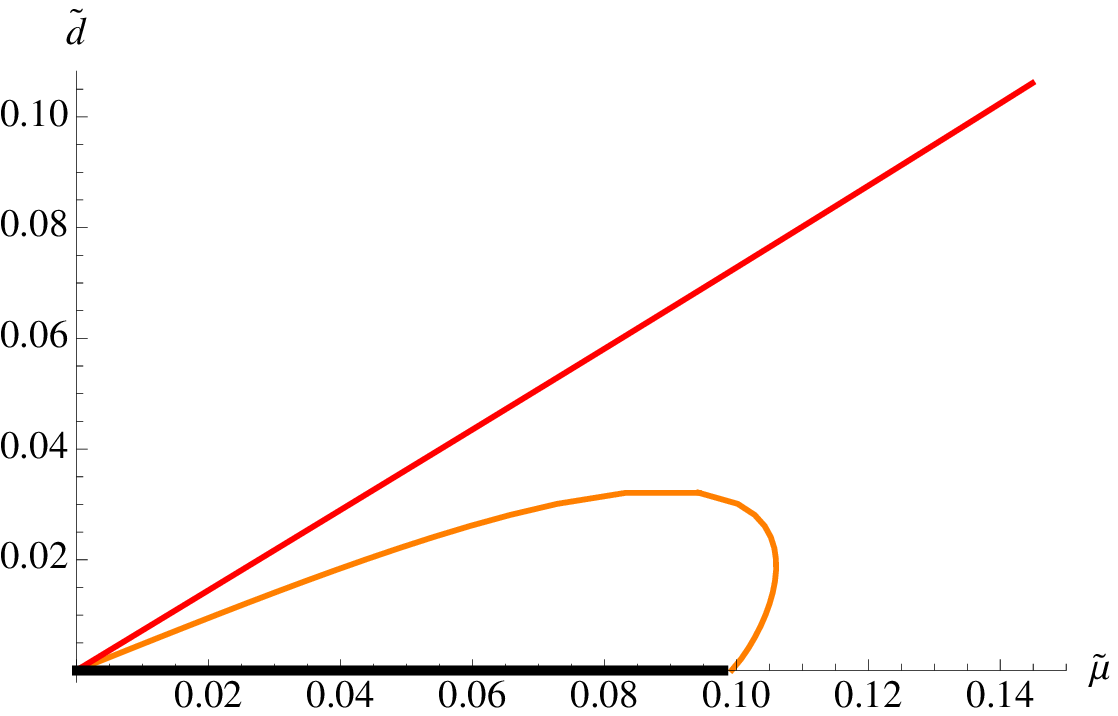}\qquad\quad
   \includegraphics[width=5cm]{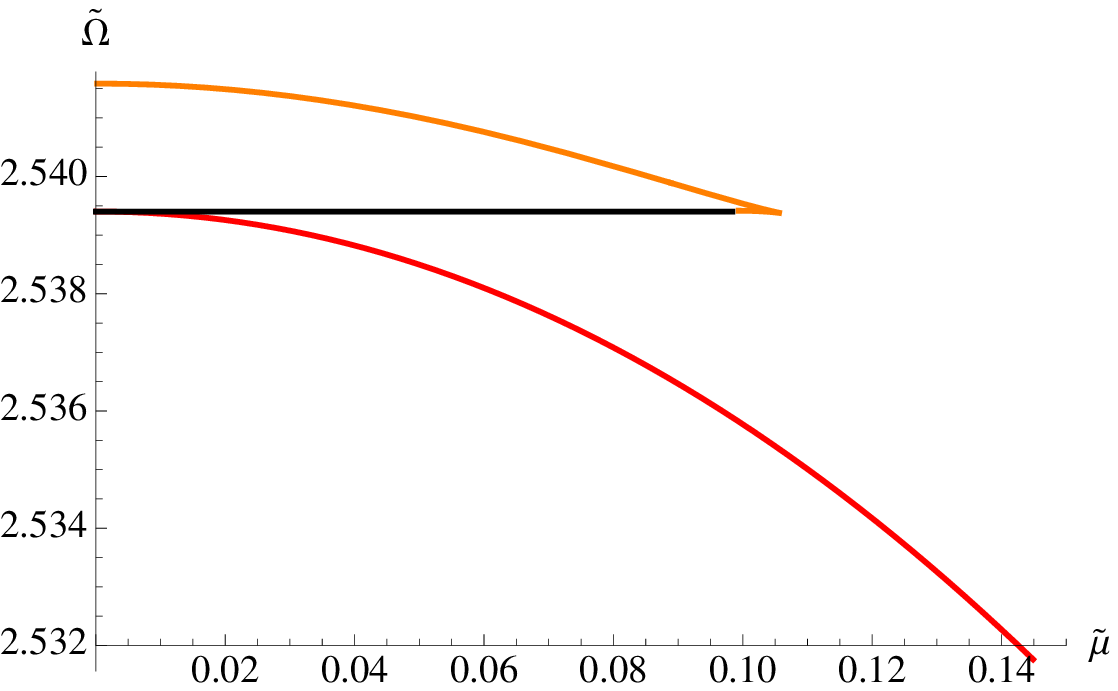}}
  \caption{ % {\Large $\!\!\!\!\!\!\!\!\;
            % \bold{\color{white}{\blacksquare}}\!\!\!\!\!\!\!\! $ }
            % 1a:
           {\small
            The D7 brane embeddings (Left), their corresponding
           $ \td - \tmu$ diagrams (Middle), and the grand potentials (Right)
           for massless quarks in the presence of a magnetic field at a variety of temperatures
           that represent slices through the phase diagram Fig \ref{Tvsmu}.
           (Parameters are scaled or $B=1/2R^2$
           in terms of parameters without tilde.)}
           }\label{Fig3}
\end{figure*}

First of all it will be interesting to see how the repulsion from
the origin induced by a magnetic field and the attraction to the
origin by the density compete. Thus we start with the canonical
ensemble (that is solutions with non-zero $\td$) and consider
black hole like embeddings exclusively. The plot in Fig
\ref{Fig2}a (Left) shows the embeddings for a small value of
density. The solutions show the chiral symmetry breaking behaviour
induced by the magnetic field but then spike to the origin by the
density at small $\trho$. For $\tm=0$ one should compare the blue
and red embeddings - the blue one is thermodynamically preferred
as shown in Fig \ref{Fig2}a (Right) The theory shows similar
behavior to that seen at zero density: there is a spiral
 structure in the $\tm$ vs $\tc$ plane (Fig \ref{Fig2}a (Middle))
 ~\cite{Johnson1}.
 That will disappear as the density increases.

As the density increases the value of the condensate for the
$\tm=0$ embeddings falls - we show a sequence of plots for growing
$\tilde{d}$ in Fig \ref{Fig2}b-d (Middle).
There is a critical value of
$\td=0.3197$ where $\tc$ becomes zero for the massless embeddings -
above this value of $\td$, D7 embedding is flat and lies along
the $\tilde{\rho}$ axis (\ref{Fig2}c-d (Left)) .
One can see from the plots that there is
a second order phase transition to a phase with no chiral
condensate. In Fig \ref{Fig2}c (Left) and \ref{Fig2}d (Left)
 we show embeddings at the critical
value of $\td$ and above it respectively. At very large density
the solutions become the usual spike embeddings of the ${\cal
N}=2$ theory at zero magnetic field.

\begin{figure}[!t]
\centering
 \subfigure[{\small $\,$ Density: the solid lines are calculated from
    (\ref{mu}) and the dotted lines are obtained by numerically
    differentiating the grand potential
    ($\td = - \frac{\del \wt{\Omega}}{\del \tmu}$). }]
  {\includegraphics[width=6cm]{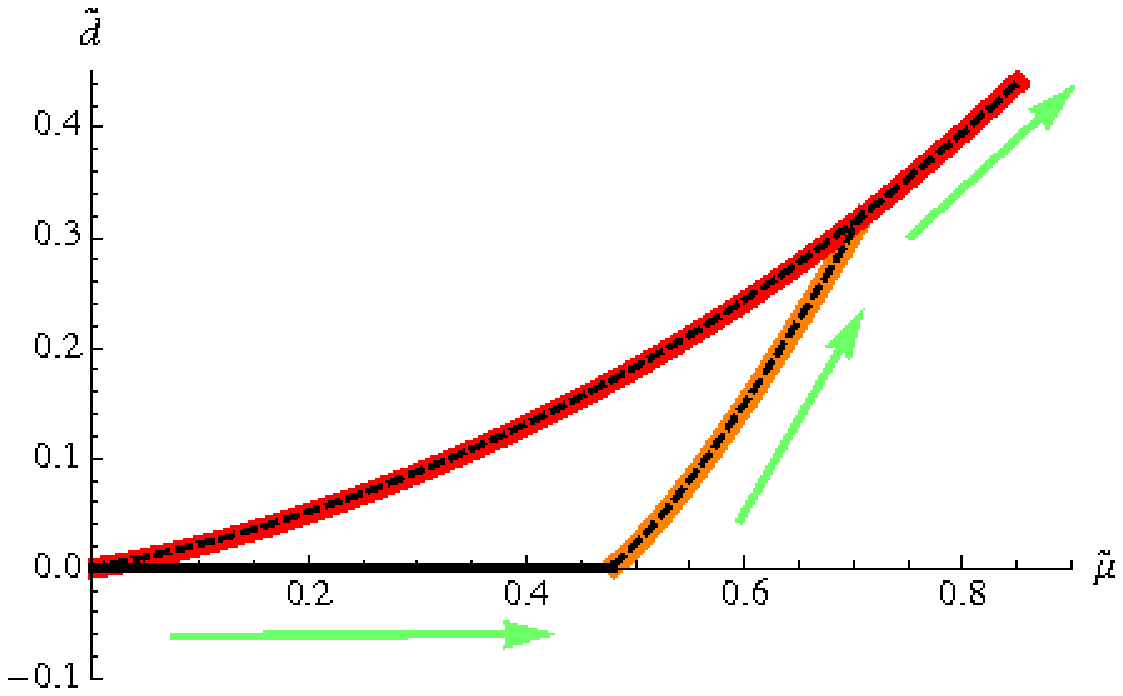}}
  \subfigure[{\small $\,$ The quark condensate.}]
  {\includegraphics[width=6cm]{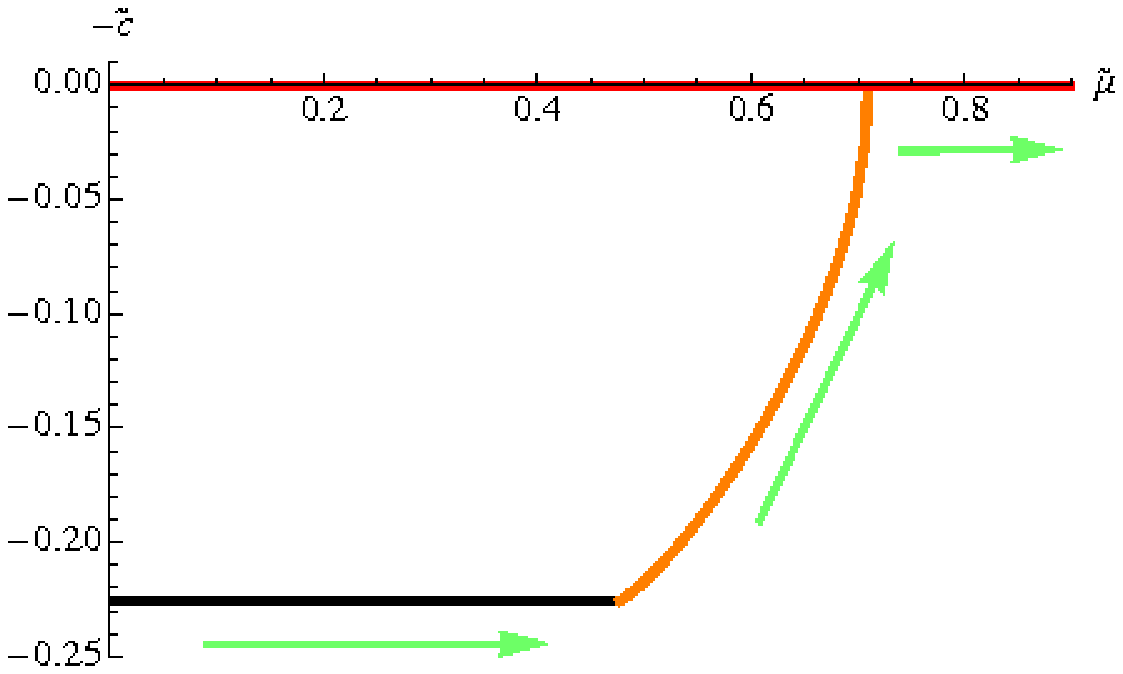}}
  \caption{
           {\small Plots of the order parameters vs chemical potential
               at zero temperature and finite $B$. Both are continuous
               across the Minkowski to spiky embedding transition($\tmu \sim 0.47$).
               The green arrows indicate the changes of phase.
          }
           }\label{Fig3sub1}
\end{figure}

We are not yet done though since there are also Minkowski
embedding with zero density but constant chemical potential.   
These can have lower energy and be the preferred
vacuum at a given value of chemical potential - that is, they are
important in the Grand Canonical Ensemble. The relevant analysis
is in Fig \ref{Fig3}a (Fig \ref{Fig3}b-e will be explained 
in the next section).  On the left it shows the three possible
types of embedding of the D7 for a given chemical potential at
zero temperature.  The black curve is the Minkowski embedding
(with $\td=0$), the blue the chiral symmetry breaking spike
embedding (with $\td \neq0$) and the red the chiral symmetry
preserving black hole embedding (with $\td \neq 0$). Strictly
speaking there is a fourth embedding which lies along the
$\tilde{\rho}$ axis and has constant $A_t=\mu$ - its energy is
equal for all $\tilde{\mu}$ to that of the red embedding at
$\tilde{\mu}=0$ and is never preferred over the red embedding with
density, so we will ignore it hence forth. The trajectory of the
three key embeddings in the $\td-\tilde{\mu}$ space is shown in
the middle plot (note that again these two variables are
thermodynamical conjugate variables). Finally on the right the
grand potential is computed. Clearly at low chemical potential the
Minkowski embedding is preferred and $\td=0$. There is a critical
value of $\tilde{\mu}=0.470$ at which a  transition occurs to the
spike embedding. This transition looks naively first order since
it is a transition between a Minkowski embedding and a black hole
embedding. However, we can see that the Grand Potential appears
smooth and the quark density is continuous, which is shown again
in Fig.4a. The solid lines in Fig \ref{Fig3sub1}a are calculated
from (\ref{mu}), which is based on the holographic dictionary. The
dotted lines are obtained by numerically differentiating the grand
potential ($\td = - \frac{\del \wt{\Omega}}{\del \tmu}$), which
comes from a thermodynamic relation. This is a nontrivial
consistency check of the holographic thermodynamics as well as our
calculation~\cite{Kim,Nakamura}.

Further in Fig \ref{Fig3sub1}b we plot the behaviour of the quark
condensate through this transition. The density and quark
condensate are both smooth and the transition looks clearly second
order. Here we have tested the smoothness numerically at better
than the 1\% level. Whether there is some other order parameter
that displays a discontinuity is unclear but it would be
surprising that any order parameter were smooth, were the
transition first order. We conclude the transition is second order
(or so weakly first order that it can be treated as second
order). This second order nature of the transition from a
Minkowski to a spiky embedding has been shown also in the $B=0,
\tm \ne 0$ case at zero temperature analytically~\cite{Karch1} and
numerically~\cite{Myers3}.

Finally, above the chemical potential corresponding to the meson
melting transition ($\tmu=0.470$), non-zero density is present and the
physics already described in the Canonical Ensemble occurs, which
turns out to be equivalent to the results from the current Grand
Canonical Ensemble. Both Ensemble predict the second order
transition to the flat embedding at the same point, $\tmu = 0.708$
or $\td = 0.3197$, which is the chiral symmetry restoration point.
Notice that for the Canonical Ensemble we used ($\tm$,$\tc$)
conjugate variables on constant $\td$ slices, while for the Grand
Canonical Ensemble we used ($\tmu$,$\td$) conjugate variables on
constant $\tm=0$ slices. This agreement from different approaches
is another consistency check of our calculation.

On the gauge theory side of the dual, the description is as
follows. At zero density there is a theory with chiral symmetry
breaking and bound mesons. As the chemical potential is increased
$\td$ remains zero and the quark condensate remains unchanged.
Then there is a second order transition to finite density
(to a spike like embedding) which is presumably associated with meson melting
induced by the medium.  At a higher density there is then a
further second order transition to a phase with zero quark
condensate.

Finally we note a recent paper~\cite{Seo} that proposed an alternative ground 
state for a chiral symmetry breaking theory at finite density. They
proposed that the string spike might end on a wrapped D5 brane baryon vertex
in the centre of the geometry. We have not considered that possibility
here but it might be interesting to investigate this in the future. The
magnetic field induced chiral symmetry breaking provides a system in
which this could be cleanly computed without the worries of the hard wall 
present in that geometry.

\section{The phase diagram in the
grand canonical ensemble}

\begin{figure}[!b]
\centering
  \includegraphics[width=7cm]{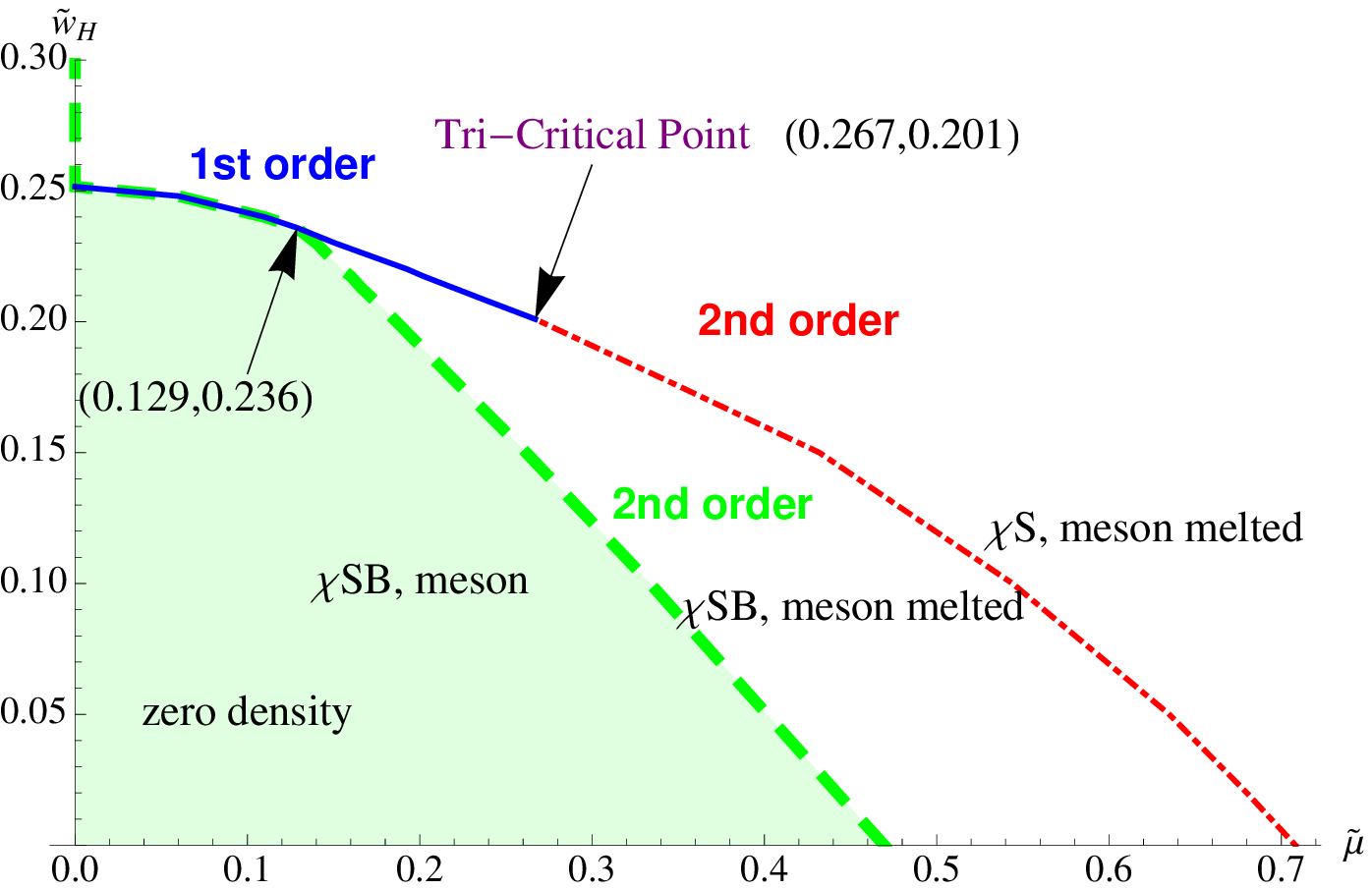}
  \caption{{\small
           The phase diagram of the ${\cal N}=2$ gauge theory with
  a magnetic field.
  The temperature is controlled by the parameter $\tilde{w}_H$
  and chemical potential by $\tmu$.
  (Parameters are scaled or $B=1/2R^2$
           in terms of parameters without tilde.)}
           }\label{Tvsmu}
\end{figure}

We have identified a first order phase transition from a chiral
symmetry breaking phase with meson bound states to a chirally
symmetric phase with melted mesons in our massless theory in the
presence of a magnetic field with increasing pure temperature. On
the finite density axis the meson melting transition is second
order and separate from another second order chiral symmetry
restoring phase transition. Clearly there must be at least one
critical point in the temperature chemical potential phase
diagram. We display the phase diagram of the massless theory,
which we will discuss the computation of, in Fig \ref{Tvsmu}.

To construct the phase diagram we have plotted slices at fixed
temperature and varying chemical potential. We display the results
in Fig \ref{Fig3}a-e where we show the embeddings (Left) relevant at different
temperatures, their trajectories in the $\td - \tmu$ plane (Middle) and the
grand potential (Right).

The phase diagram agrees with our previous results: At zero
chemical potential we have the transition point $\tw_H = 0.2516$.
At zero temperature we have the transition point at $\tmu =
0.708$, which corresponds to $\td=0.3197$. We also identify
$\tmu=0.470$ as the position of the second order transition to a
meson melted phase with non-zero $\td$ and chiral condensate
$\tc$.

\begin{figure}[!t]
\centering
 \subfigure[{\small $\, \tw_H =0.15$   }]
  {\includegraphics[width=4cm]{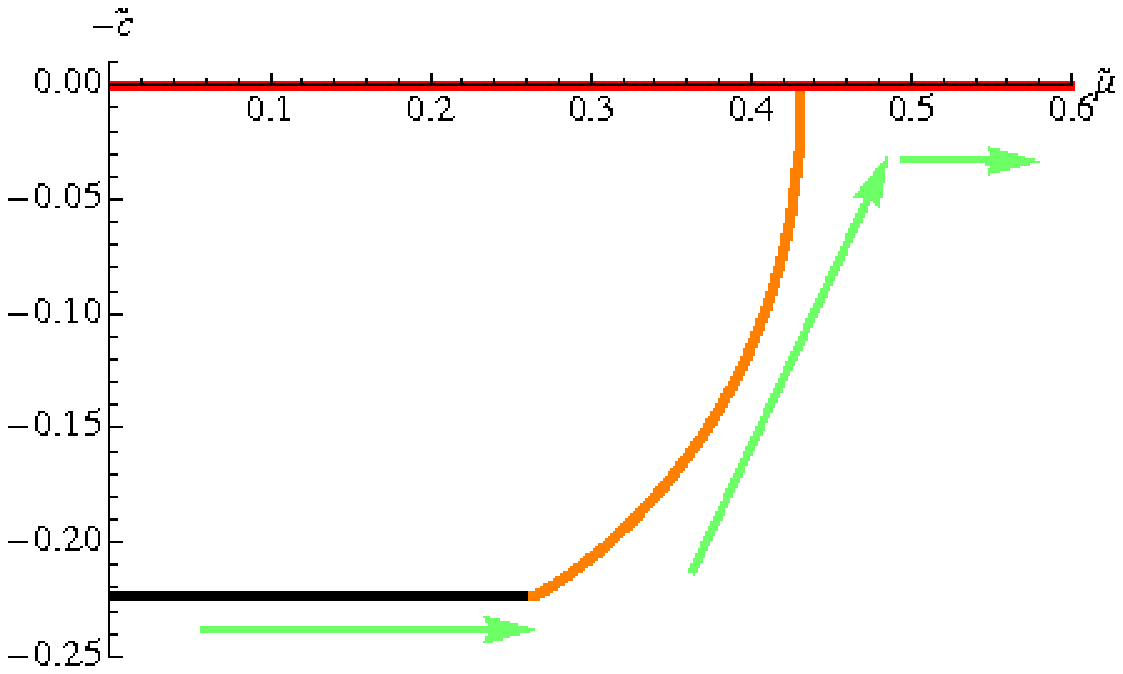}}
  \subfigure[{\small $\, \tw_H = 0.23$ }]
  {\includegraphics[width=4cm]{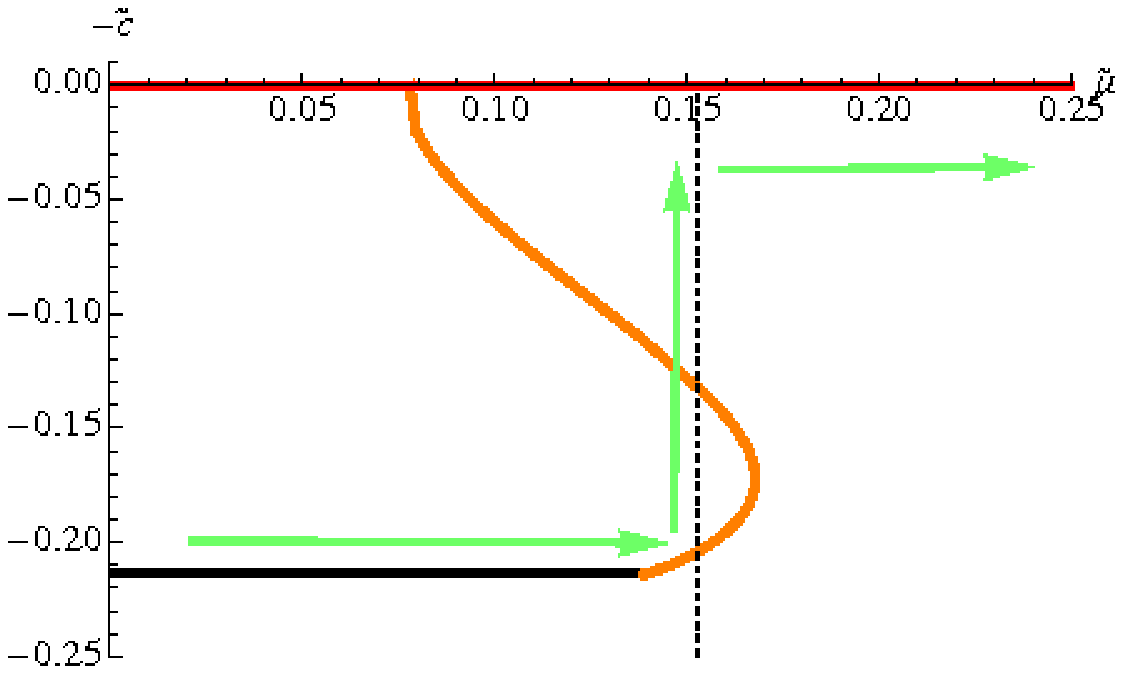}}
  \caption{
           {\small Quark condensate vs chemical potential
              at finite $B$. Both are continuous
               across the Minkowski (black) to black hole (orange) embedding transition.
               At $\tw_H = 0.23$ the black hole (orange) to black hole(red)
              transition is discontinous.  }
           }\label{Fig3sub2}
\end{figure}

The dotted green line is the line along which $\td=0$ and
corresponds to the second order meson melting transition from a
Minkowski embedding to a black hole embedding. The transition
generates density continuously from zero. The quark condensate
also smoothly decreases from its constant value on the Minkowski
embedding. We display the continuous behaviour of the quark
condensate across the transition in Fig \ref{Fig3sub2}. Note this means that
the slope of the embedding at the UV boundary is continuous
through the transition even though the embedding in the IR is
discontinuous and topology changing. Again we have checked the
smoothness of these parameters numerically to better than the 1\%
level.

\begin{figure}[!b]
\centering
  \includegraphics[width=7cm]{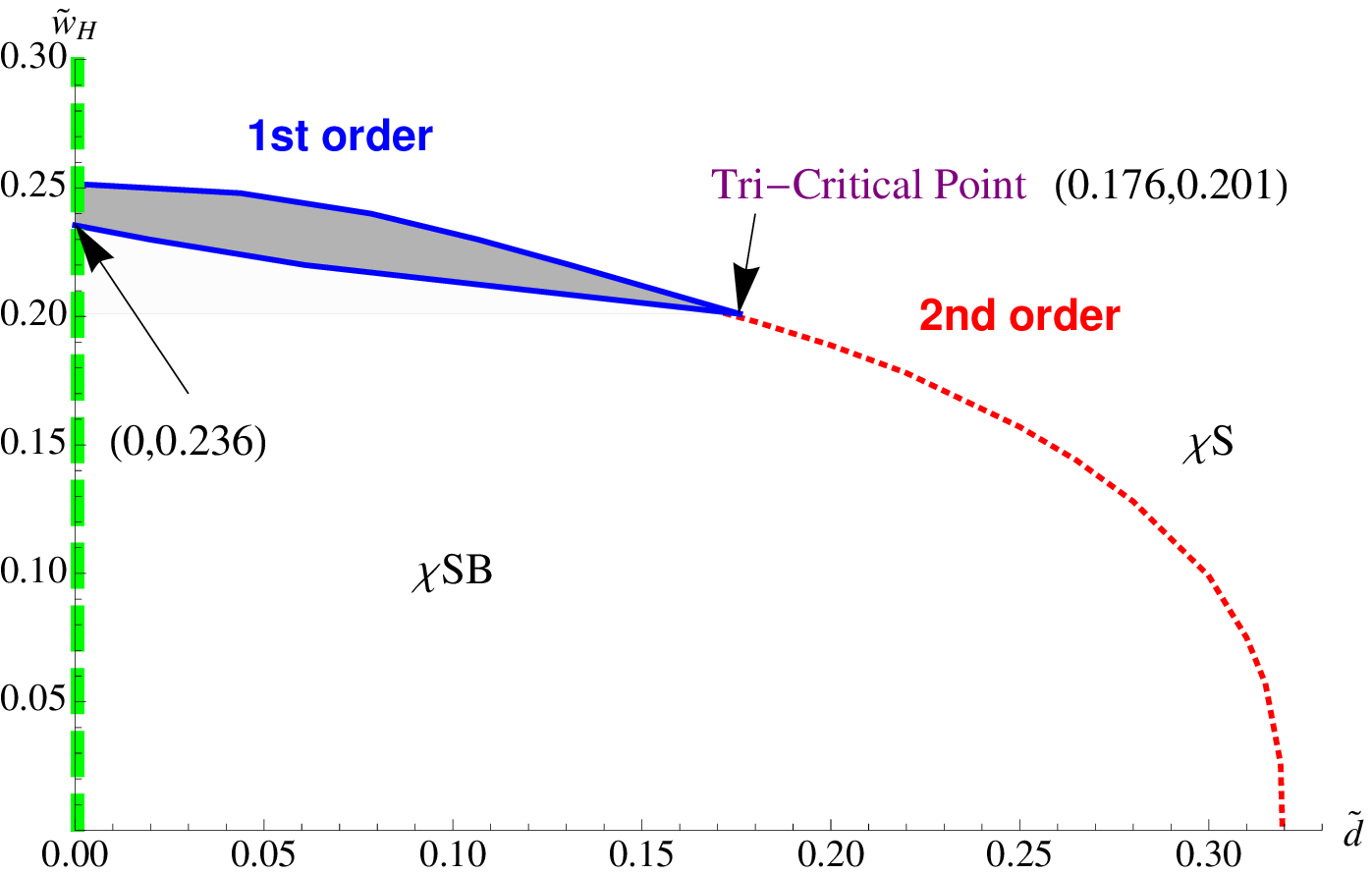}
  \caption{{\small
           The phase diagram of the ${\cal N}=2$ gauge theory with
  a magnetic field.
  The temperature is controlled by the parameter $\tilde{w}_H$
  and the density  by $\td$.
  (Parameters are scaled or $B=1/2R^2$
           in terms of parameters without tilde.)}
           }\label{Tvsd}
\end{figure}

The blue line corresponds to a first order transition and the
red dotted line is a second order transition in density, chiral
condensate etc. The red dotted line is rather special in that this
is a phase boundary only at $\tm=0$. This is because this phase
boundary is related to the spontaneous breaking of chiral symmetry
which only exists at $\tm=0$. At finite $\tm$ it must be a cross
over region as we will discuss further in section \ref{FMsection}

The diagram then displays two tri-critical points. It is
straightforward to identify where the points lie numerically. The
chiral symmetry tri-critical point where the first and second
order chiral symmetry restoration transitions join lies at the
point $(\tmu,\tw_H) = (0.267,0.201)$. The second  tri-critial
point where the meson melting transitions join is at $(\tmu,\tw_H)
=(0.129,0.236)$.

 \section{The Phase Diagram in the canonical ensemble}

We can study the phase diagram also in the canonical ensemble. It
is shown in Fig \ref{Tvsd} and has the same information as Fig \ref{Tvsmu}.
The pale green region in Fig \ref{Tvsmu} lies in the green dotted line
along the
$\tilde{w}_H$ axis of Fig \ref{Tvsd}. The chiral symmetry breaking region
enclosed by the red, green and blue lines in each figure map onto
each other. Similarly the high temperature and density region to
the upper right of all the lines in both plots map onto each
other. The two double blue lines and the area between them in Fig
\ref{Tvsd}  correspond to the single blue line in Fig \ref{Tvsmu},
which is natural
since the blue line in Fig \ref{Tvsmu} is a first order transition line and
the density change is discontinuous. Thus the gray region in Fig
\ref{Tvsd}  is an unstable density region which hides in the phase boundary
in  Fig \ref{Tvsmu}. That region may only be reached by super-cooling or
super-heating since it is unstable. The true ground state at those
densities and temperatures should be a mixture of the black hole
and Minkowski embedding in analogy with the liquid-gas mixture
between the phase transition's of water~\cite{Myers3}.
It's not clear how to
realize that mixture in a holographic set-up.

\section{Finite mass}\label{FMsection}

\begin{figure}[!t]
\centering
  \subfigure[
   The curves correspond to $\tm = 0(\mathrm{black}), 1(\mathrm{red}),1.5(\mathrm{orange}), 2(\mathrm{green}), 3(\mathrm{blue})$ from bottom to top. The gray line is the path of the critical points.]
  {\includegraphics[width=8cm]{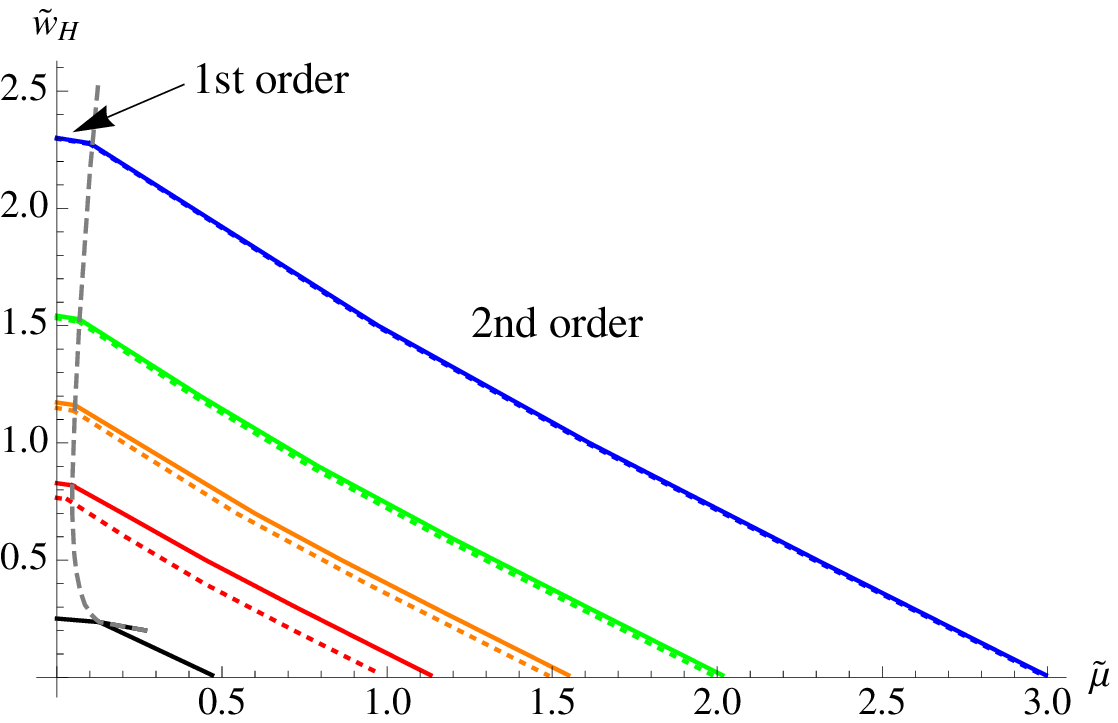}}
  \subfigure [A zoom into figure (a) to show the critical point structure.
   at small chemical potential.
  Here we show the detail of $\tm=1$ case, but a similar structure
  exists for every case.]
  {\includegraphics[width=3.8cm]{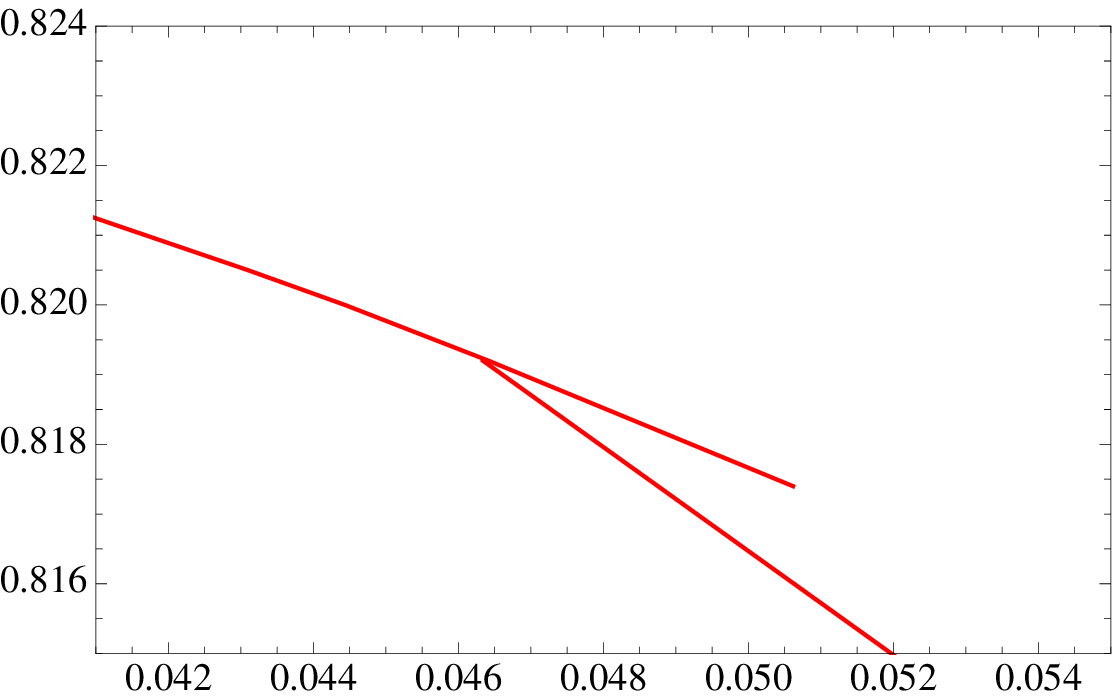}
   \includegraphics[width=4cm]{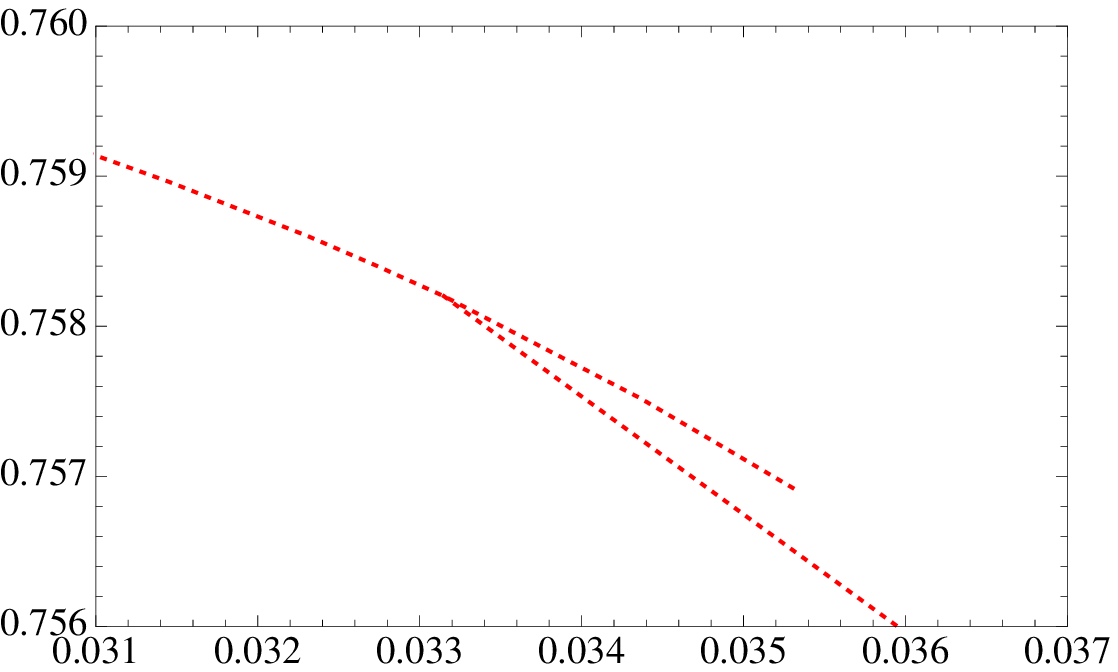} }
     \caption{
  \small{The phase diagram at finite current quark mass with
         finite B (solid lines) and zero B (dotted lines). }
} \label{FM}
\end{figure}

We next describe the evolution of the phase diagram with quark
mass. If we move away from zero quark mass then the second order
chiral symmetry restoration phase transition at T=0 but growing
chemical potential becomes a cross over transition. This can be
seen in Fig \ref{Fig2} where for $\tm \neq 0$ the non-zero value of
the condensate can be seen to change smoothly with changing $\tmu$
and there is no jump in any order parameter.
The (chiral) tri-critical point becomes a critical point. However,
the other transition lines survive the introduction of a quark
mass.

In Fig \ref{FM}. we plot the phase diagram for various quark mass, $\tm$,
at constant $B$. The colors represent different quark masses - $\tm
= 0, 1, 1.5, 2, 3$ from bottom to top  are black, red, orange,
green, and blue. The solid lines are for finite, fixed B. To show
the influence of the magnetic field we also display the $B=0$
solution as the dotted lines. The gray line shows the motion of
the critical points.

In general the magnetic field shifts the transition line up and
right, meaning that the magnetic field makes the meson more stable
against the temperature/density meson dissociation effect. This is
important at small $\tm$ but negligible at large $\tm$ as
expected.

Both critical points survive the introduction of a finite $\tm$,
even though it looks like there is no critical point in Fig \ref{FM}a.
Zooming in on the appropriate region at small chemical potential
reveals the two critical point structure as shown in Fig \ref{FM}b.
Their positions, as $\tm$ changes, are marked by the gray line in Fig
\ref{FM}a. The one line represents the two critical points which are
indistinguishable close on the scale of Fig \ref{FM}a. The chiral
symmetry critical point moves very close to the other critical
point even for a very small mass ($\tm \sim 0.01$). The
interpretation of the critical points and the phase boundaries are
the same as in the $\tm=0$ case in the previous section.

Notice that the black hole to black hole transition exists even in
the B=0 case as shown in Fig \ref{FM}b(Right), so it is not purely due to
the magnetic field.  Nevertheless this transition seems not to
have been  reported in the previous works
\cite{Myers3,Sin2}. We believe that this is because the
 transition line between the two critical points is too small to
be resolved on the scale of Fig \ref{FM}a, which  agrees qualitatively
with the figures in \cite{Myers3}. In order to find
 those transitions we had to slice the temperature down to order
$10^{-3}$ as shown on the vertical axis in Fig \ref{FM}b(Right). Any
coarser graining would miss it.

\begin{figure*}[]
\centering
 \subfigure[$\ w_H = 0.4$.  At $\bar{\mu} \sim 0.4467$ there is a Minkowski to black hole embedding transition, which is second order in both chiral condensation and density.]
  {\includegraphics[width=5cm]{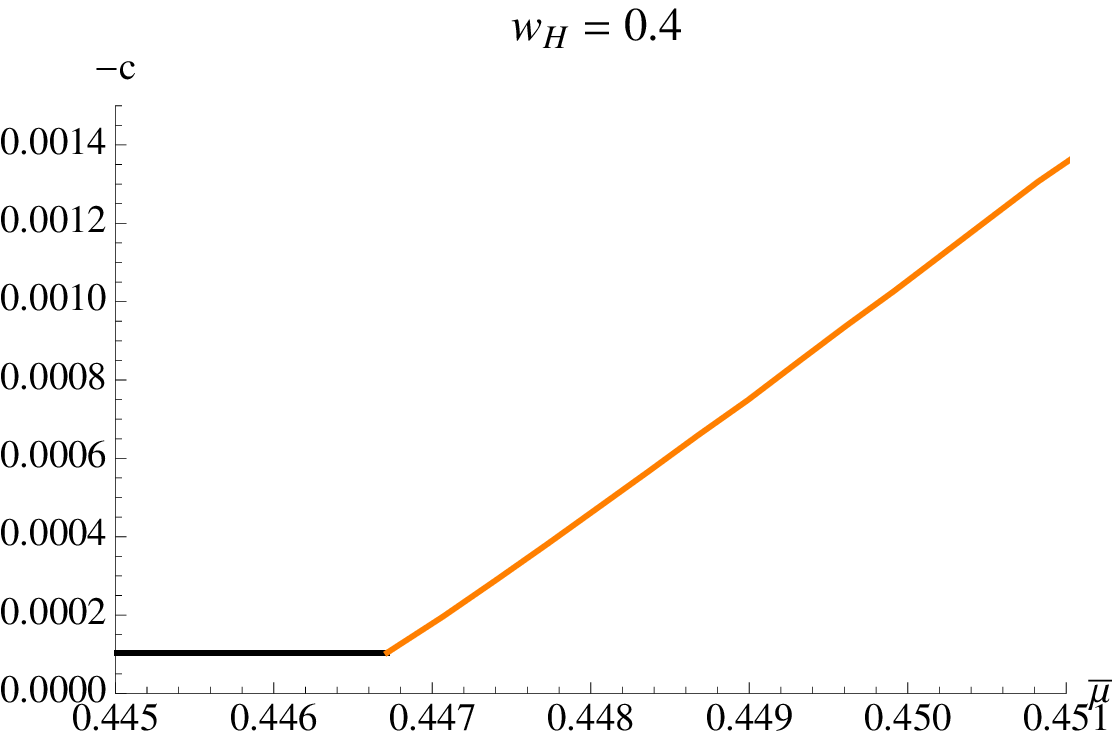}\qquad
   \includegraphics[width=5cm]{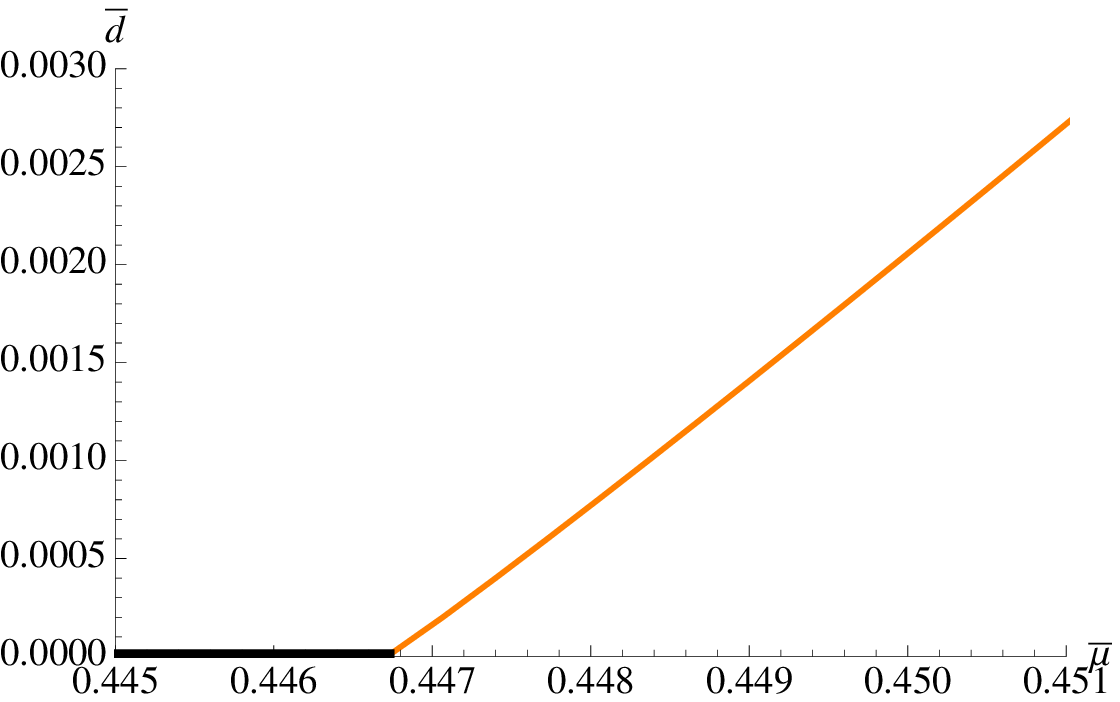}\qquad\quad
   \includegraphics[width=5cm]{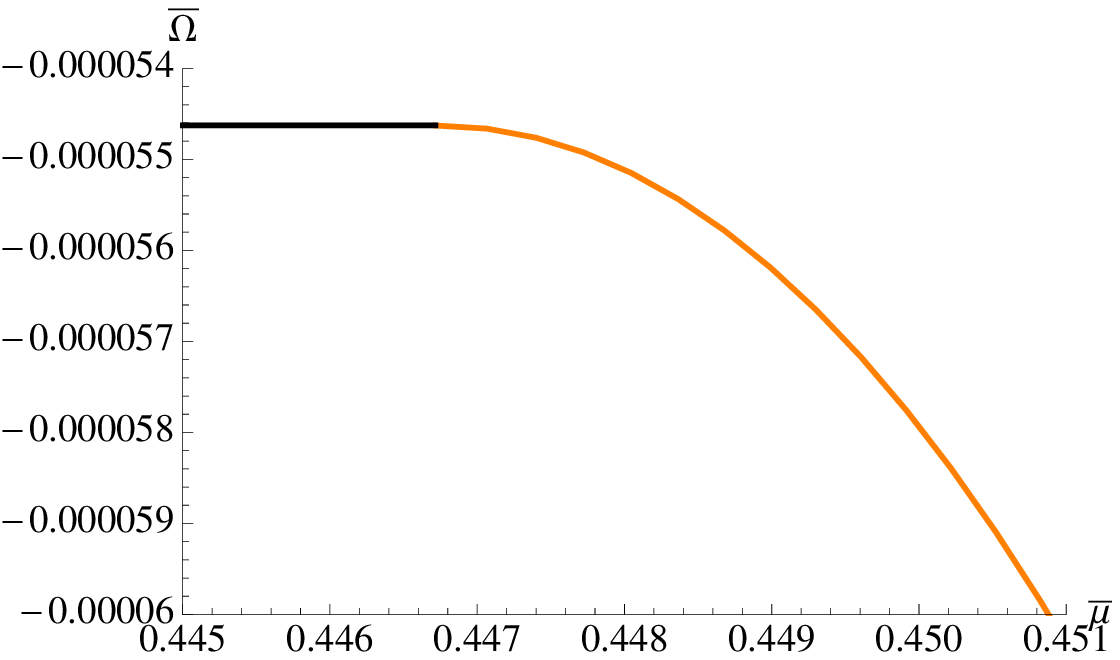}}
  \subfigure[$\ w_H = 0.7575$. There are two transitions.
  The first ($\bar{\mu} \sim 0.0341$) is a Minkowski to
  black hole transition and second order in condensation and density.
  The second ($\bar{\mu} \sim 0.03455$) is a black hole to black hole transition
  and first order. ]
  {\includegraphics[width=5cm]{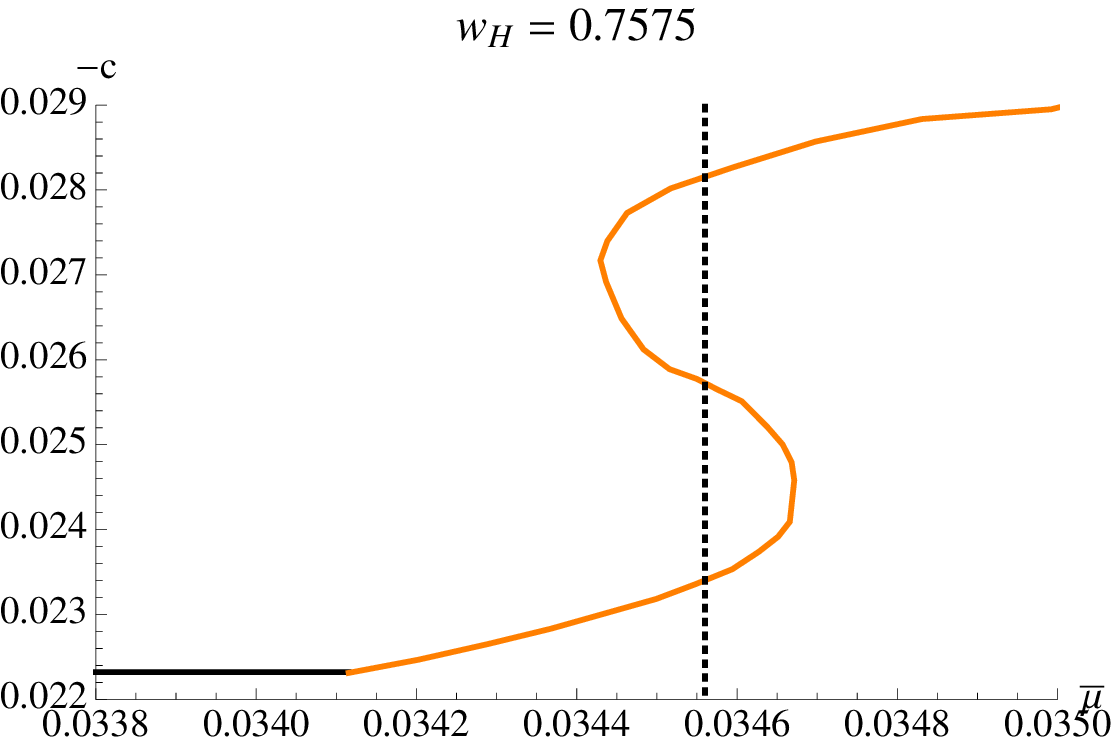}\qquad
   \includegraphics[width=5cm]{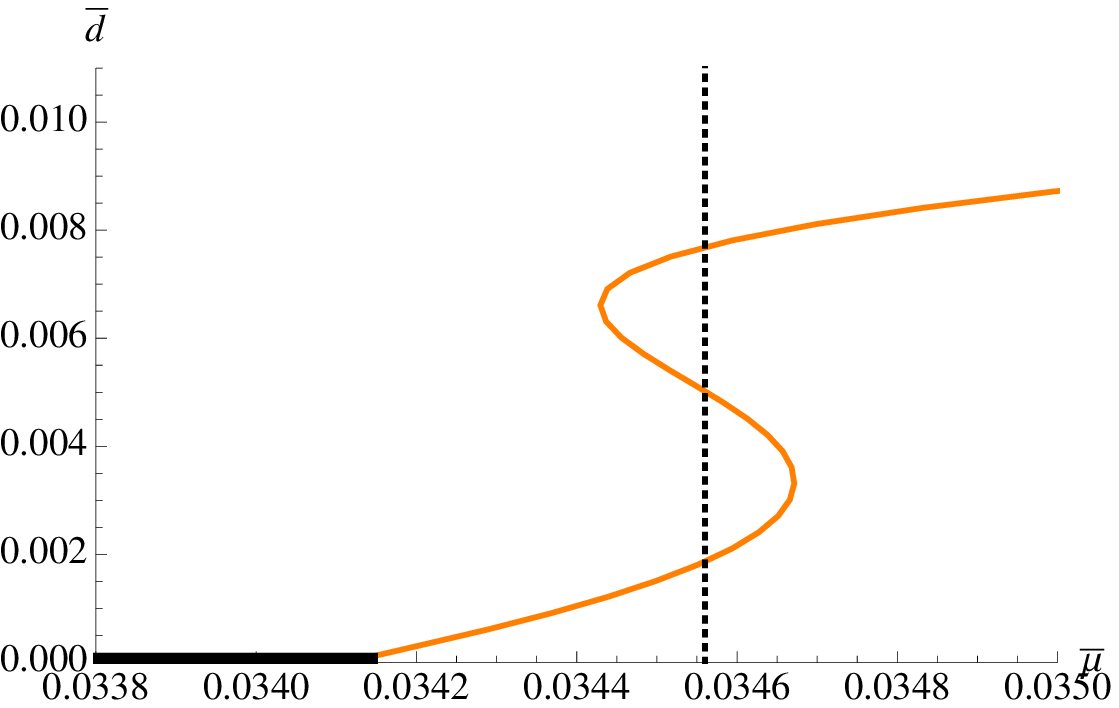}\qquad\quad
   \includegraphics[width=5cm]{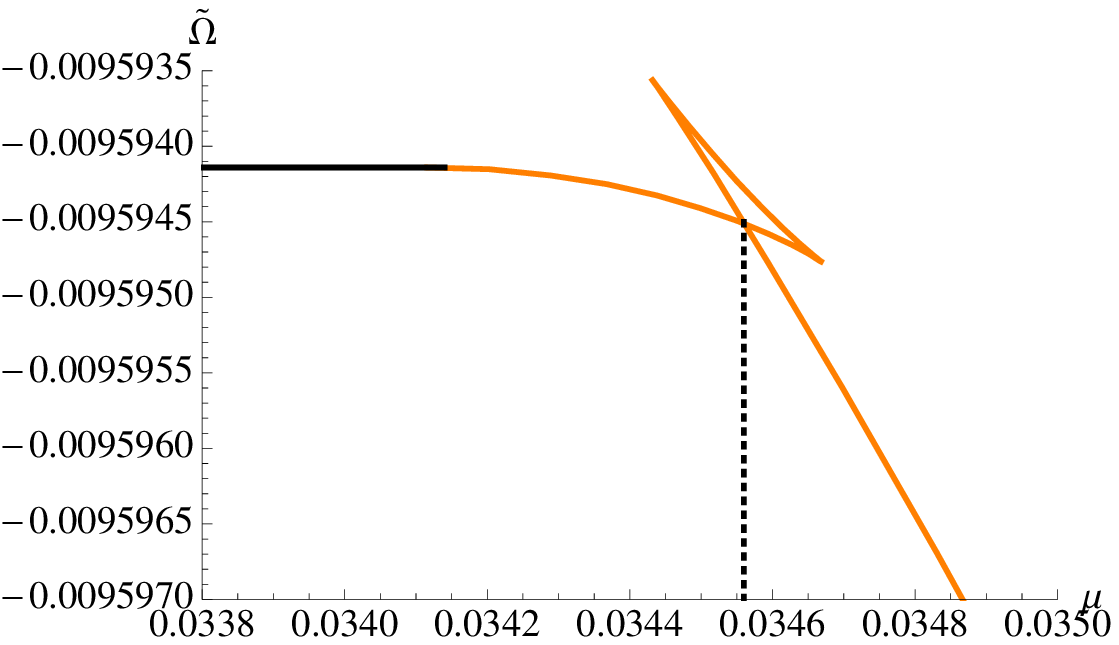}}
  \subfigure[$\ w_H = 0.7587$.  At $\bar{\mu} \sim 0.0321$ there is a Minkowski
  to black hole embedding transition, which is first order in both chiral condensation and density.]
  {\includegraphics[width=5cm]{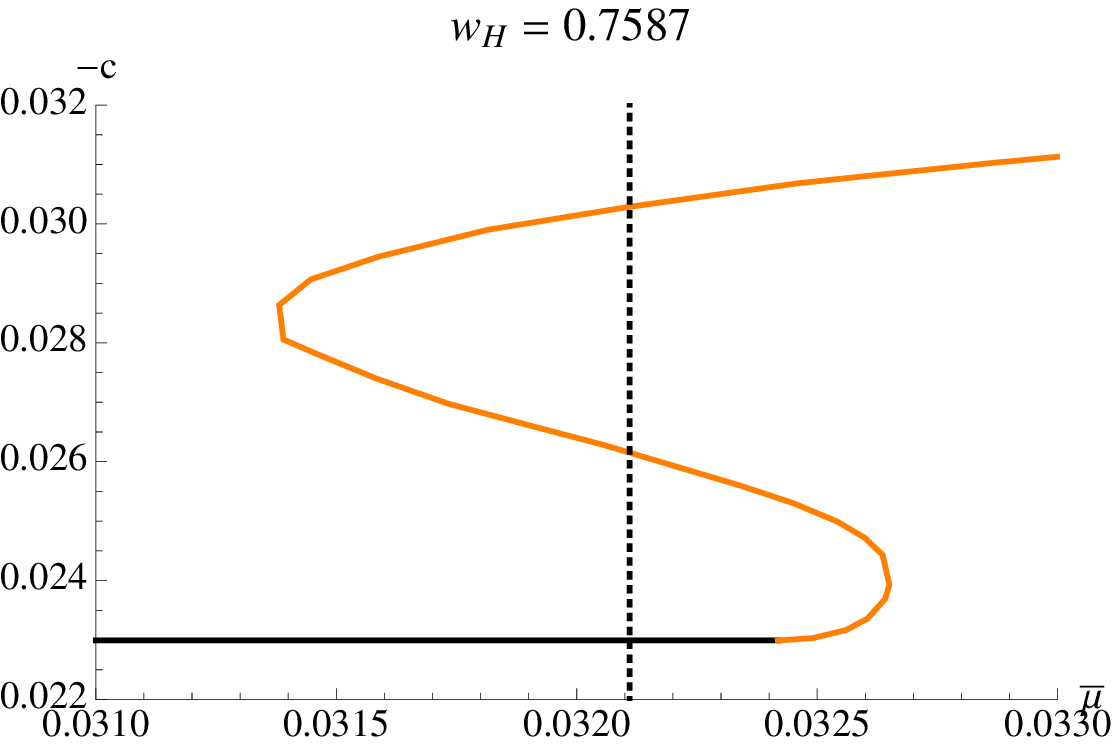}\qquad
   \includegraphics[width=5cm]{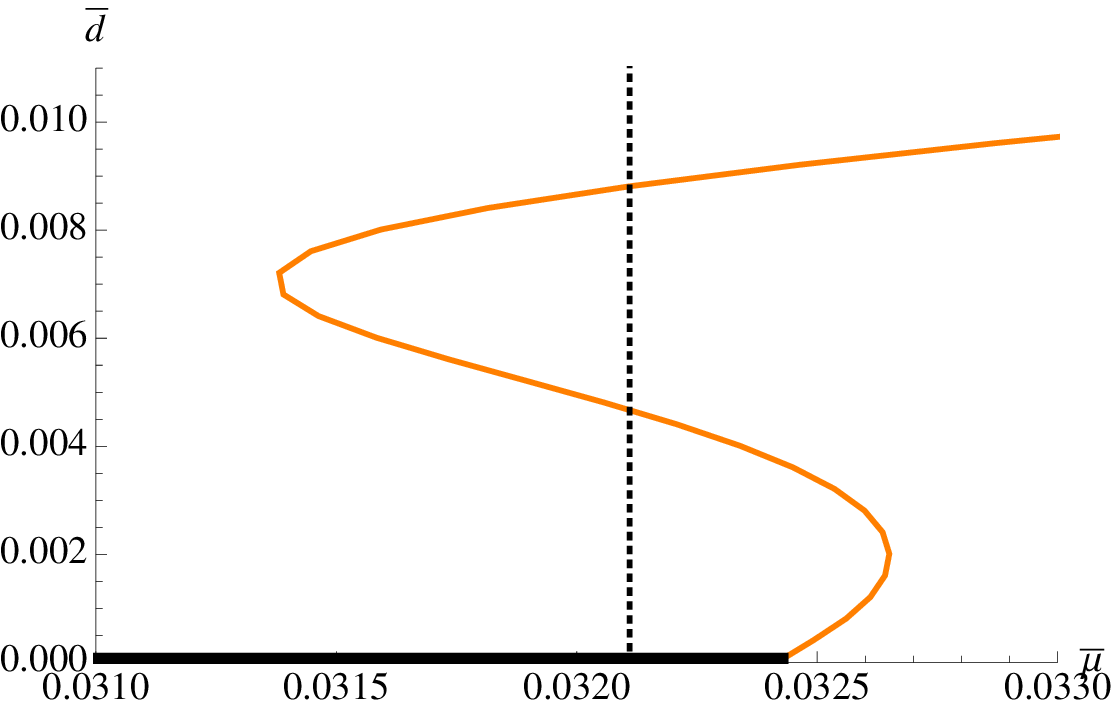}\qquad\quad
   \includegraphics[width=5cm]{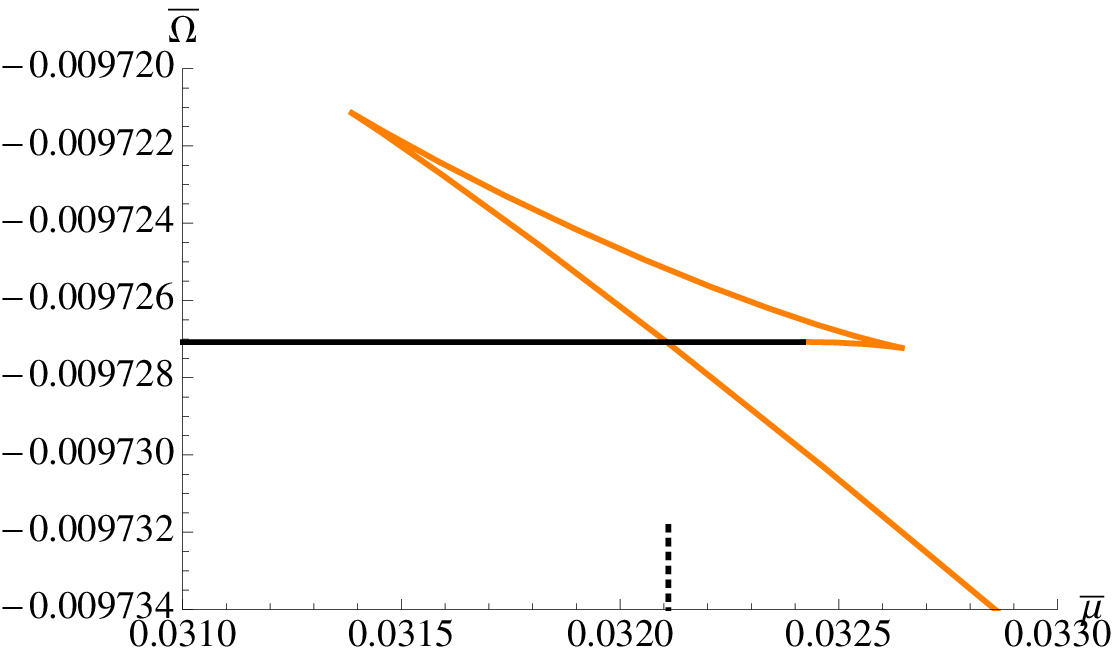}}
  \subfigure[$\ w_H = 0.762$.  At $\bar{\mu} \sim 0.0235$ there is a Minkowski
  to black hole embedding transition, which is first order in both chiral condensation and density. ]
  {\includegraphics[width=5cm]{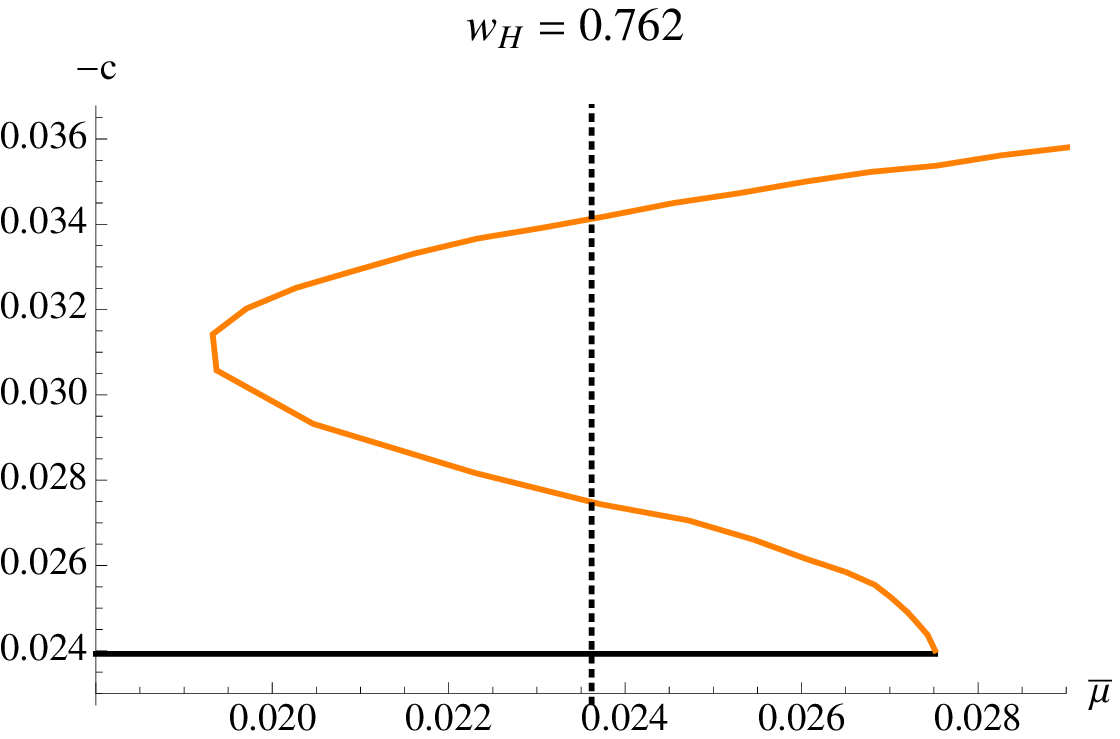}\qquad
   \includegraphics[width=5cm]{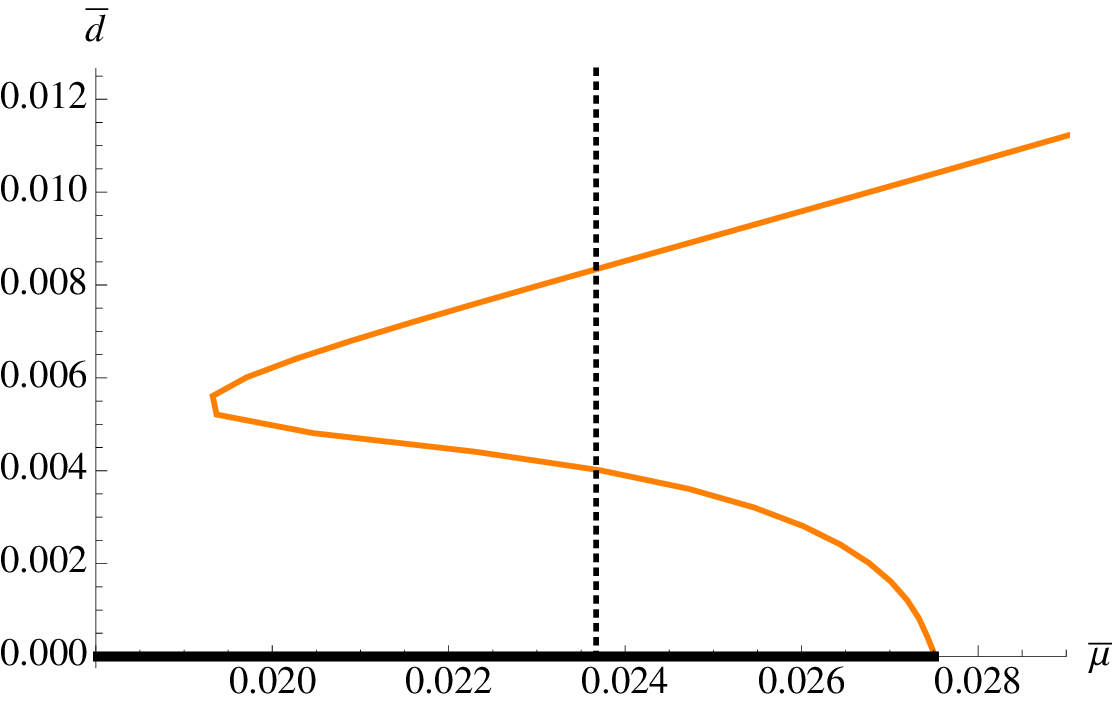}\qquad\quad
   \includegraphics[width=5cm]{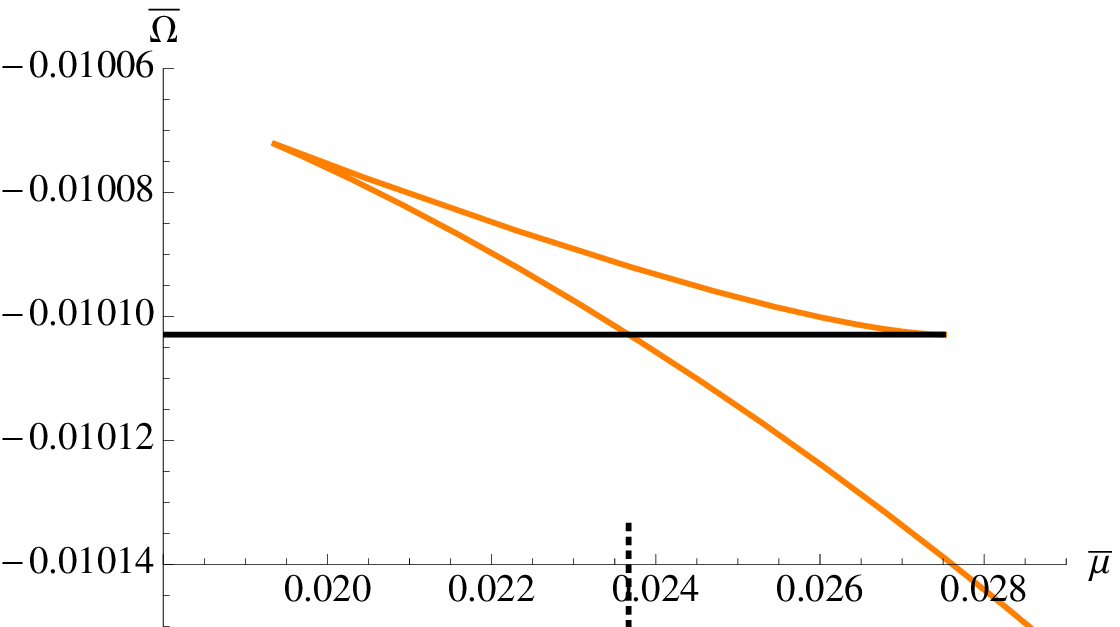}}
  \subfigure[ Above $\ w_H = 0.7658$ only a black hole embedding (Red) is stable configuration.]
  {\includegraphics[width=5cm]{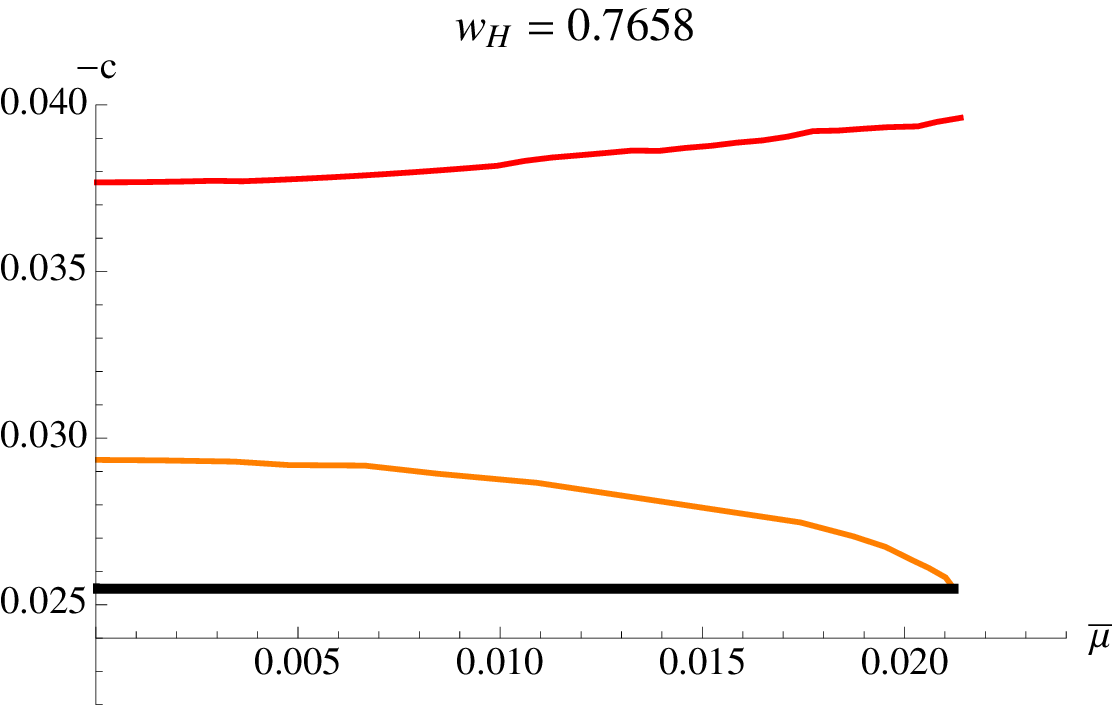}\qquad
   \includegraphics[width=5cm]{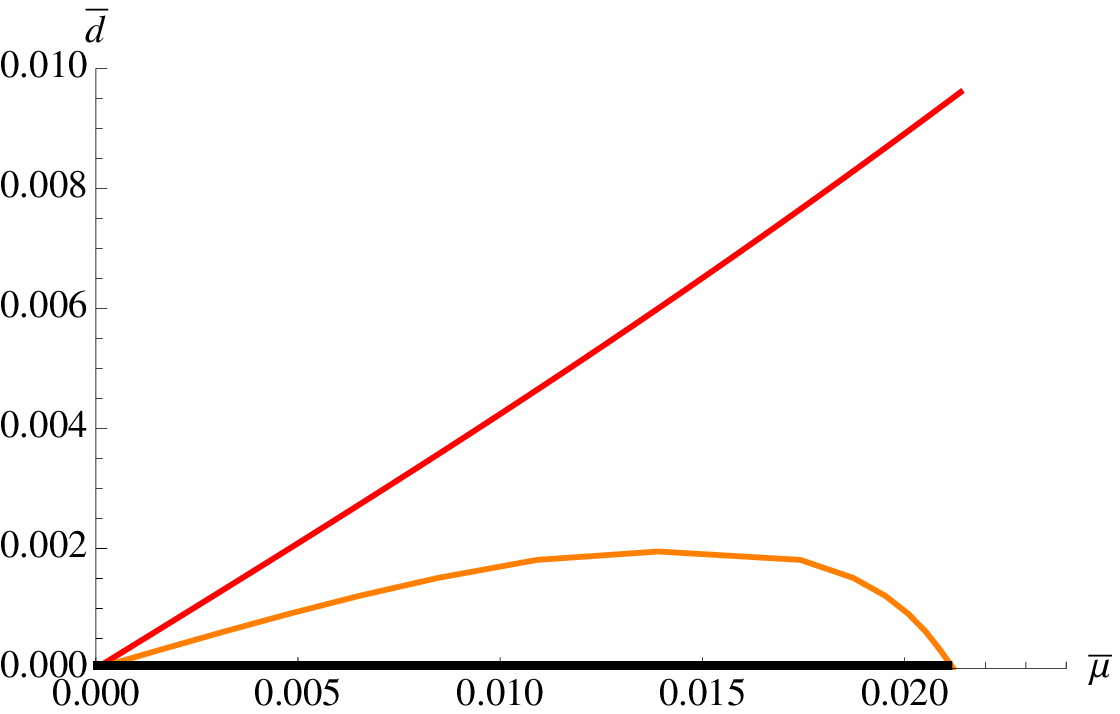}\qquad\quad
   \includegraphics[width=5cm]{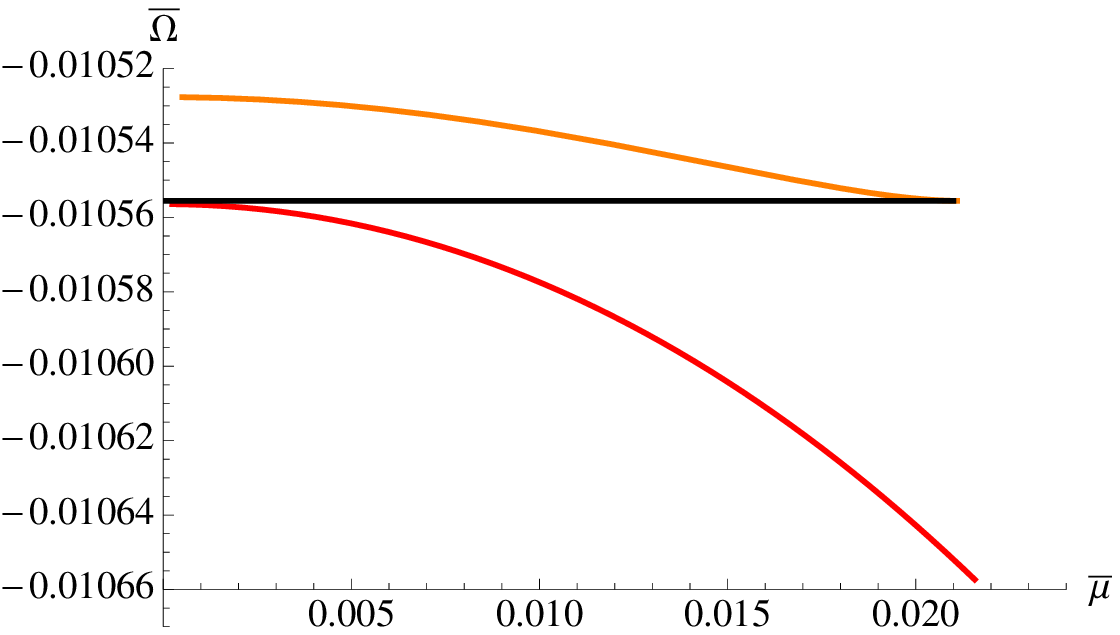}}
  \caption{ % {\Large $\!\!\!\!\!\!\!\!\;
            % \bold{\color{white}{\blacksquare}}\!\!\!\!\!\!\!\! $ }
            % 1a:
           {\small
            Chiral condensation (Left), density
           (Middle), and the grand potentials (Right)
           for massive quarks ($m=1$) at $B=0$ at a variety of temperatures
           that represent slices through the phase diagram Fig \ref{FM}a.
           }
           }\label{B0}
\end{figure*}

 The final surprise relative to the previous work is that the
meson melting transition below the critical point appears second
order in our work even in the infinite mass limit. To emphasize
this we show a number of plots in the $B=0$ theory in Fig \ref{B0}.

Since the scaled variables (\ref{scaled}) cannot be used at $B=0$,
 (\ref{Omega}) and (\ref{mu}) read in terms of the original coordinates:
\begin{eqnarray}
  && \bar{\Omega}(w_H,\bar{\mu}) :=  \frac{-{S}}{N_f T_{D7}
   \mathrm{Vol}} \nn \\
  && \quad
  = \int_{\rho_H}^\infty d\rho\  \frac{w^4 - w_H^4}{w^4 }
  \sqrt{\frac{(1+(L')^2)}{K}} \left(\frac{w^4+w_H^4}{w^4}\right)^2
   \rho^6 \ , \nn \\ \label{Omega1}
\end{eqnarray}
where
\begin{eqnarray}
  \bar{\mu}
  &=& \int_{\rho_H}^\infty d\rho\ \bar{d}\ \frac{w^4 - w_H^4}{w^4 + w_H^4} \sqrt{\frac{ 1+(L')^2 }{K}} \ , \label{mu1} \\
   K &=& \left(\frac{w^4+w_H^4}{w^4}\right)^2 \rho^6
              + \frac{w^4}{(w^4+w^4_H)} \bar{d}^2 \ , \\
  \bar{\mu} &:=& \sqrt{2}^3\pi\a'A_t(\infty) \ , \quad
  \bar{d} := \frac{\sqrt{2}^3}{N_f T_{D7} 2\pi \a'} d
\end{eqnarray}
By the same procedures as in the previous sections we get Fig \ref{B0}.
Compared to Fig \ref{Fig3}, the left column of Fig \ref{B0} is the chiral
condensate instead of the embedding configurations. In Fig \ref{Fig3}
there is always a red black hole embedding, which corresponds to the
flat embedding at zero quark mass. It is not present at finite
quark mass.

At very low temperature the transition is Minkowski to black hole
and {\it second order} in the condensate and density(Fig \ref{B0}a). As
the temperature goes up a {\it new} black hole to black hole
transition pops up by developing a `swallow tail' in the grand
potential - this transition is first order in the condensate and
density(Fig \ref{B0}b). As temperature rises the `swallow tail' grows
continually and eventually ``swallows" the {\it second order}
Minkowski to black hole transition (Fig \ref{B0} c,d). \ie $ $ At higher
temperature the {\it second order} Minkowski to black hole
transition enters an unstable regime and plays no role any more.
Instead only the {\it first order} Minkowski to black hole
transition is manifest. Finally the Minkowski embedding becomes
unstable compared to the black hole embedding(Fig \ref{B0}e). At an even
higher temperature the Minkowski embedding is not allowed and only
a black hole  embedding is available (Not shown in Fig \ref{B0}).

These results all match with our work at finite $B$ and
increasing mass, confirming those results and our phase diagrams
already presented.

\section{Comparison to QCD}

\begin{figure}[!t]
\centering
 \subfigure[{\small Standard scenario}]
  {\includegraphics[width=4cm]{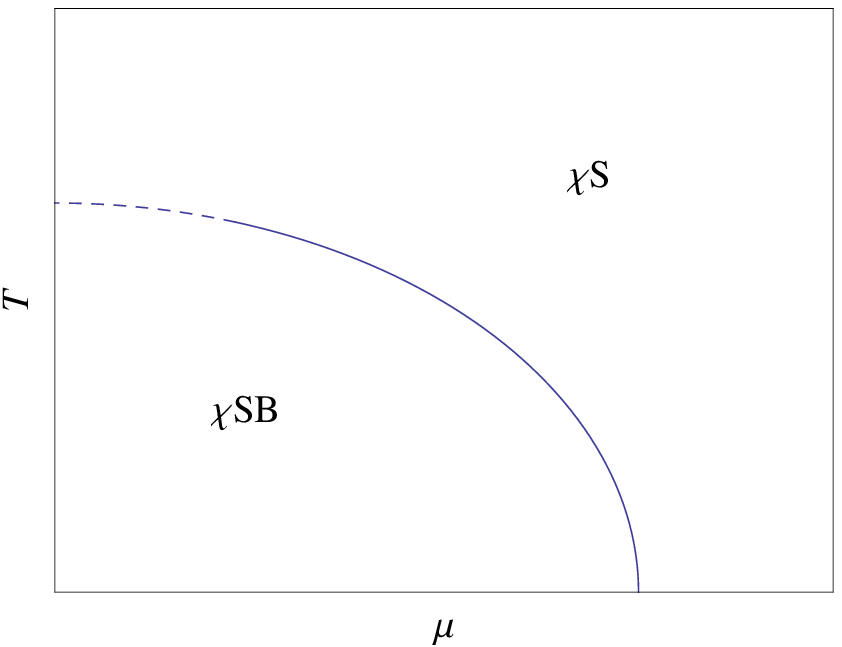}}
  \subfigure[{\small Exotic Scenario}]
  {\includegraphics[width=4cm]{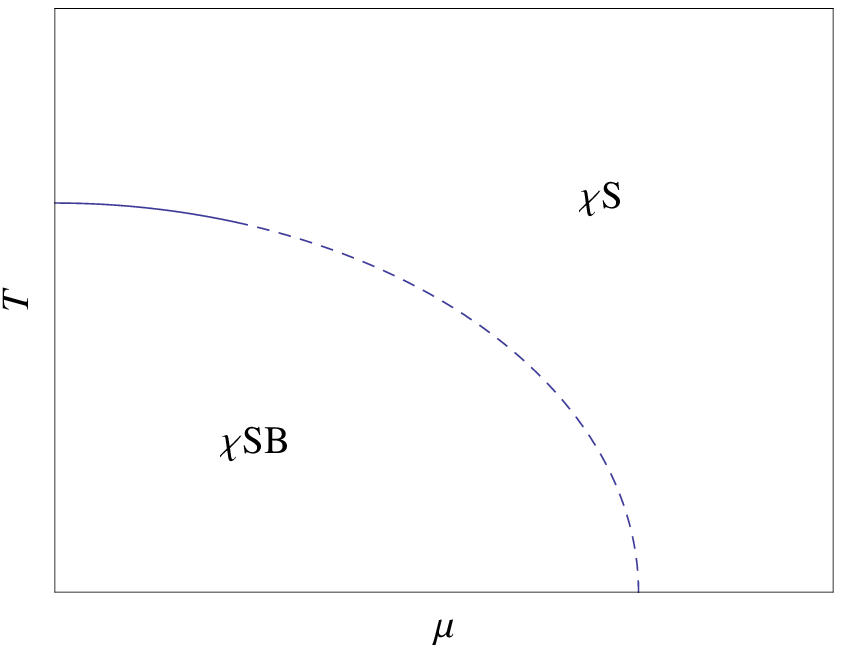}}
  \caption{
           {\small Two possible phase diagrams for QCD with the observed quark masses.
           (a) is the standard scenario found in most of the literature but a diagram
           as different as (b) remains potentially possible according to the work in \cite{Owe}.
           We have not included any color superconducting phase here at large chemical potential. }
           }\label{QCD}
\end{figure}

We have computed the phase diagram for a particular gauge theory
using holographic techniques. There are many differences between
our theory and QCD: the theory has super partners of the quarks
and glue present; it is at large $N$ and small $N_f$, so quenched
(and we have only computed for degenerate quarks to avoid
complications involving the non-abelian DBI action); the theory
has deconfined glue for all non-zero temperature; the theory has a
distinct meson melting transition. In spite of these differences
the phase diagram for the chiral condensate shows many of the
aspects of the QCD phase diagram so we will briefly make a
comparison here.

The QCD phase diagram is in fact not perfectly mapped out since
there have only recently been lattice computations attempting to
address finite density \cite{Owe}.  The phase structure also
depends on the relative masses of the up, down and strange quarks.
The standard theoretical picture
\cite{Rajagopal:2000wf,Stephanov:2007fk,Owe} for physical QCD is
shown in Fig \ref{QCD}a. At zero chemical potential
the transition with temperature
is second order (or a cross over with massive quarks). At zero
temperature there is a first order transition with increasing
chemical potential (ignoring any superconducting phase). These transitions
are joined by a critical point. Comparing to our theory in Fig \ref{Tvsmu}
we see that the transitions' orders are reversed and the pictures
look rather different.

In fact though as argued in \cite{Owe} the picture could be very
different in QCD. At zero quark mass the finite temperature
transition is first order and whether it has changed to second
order depends crucially on the precise physical quark masses.
Similarly whether the finite density transition is truly first
order or second order depends on the exact physical point in the
$m_{u,d}$, $m_s$, $\mu$, T volume. Arguments can even be made for
a phase diagram matching that in Fig \ref{QCD}b which then matches the
structure of the chiral symmetry restoring phase diagram of the
theory we have studied. For the true answer in QCD we must wait on
lattice developments. Clearly our model will not match QCD's phase
diagram point by point in $m_{u,d}$, $m_s$, $\mu$, T volume but it
provides an environment in which clear computation is possible for
structures that match some points in that phase space.

Finally, we note a more general point that seems to emerge from the
analysis. The introduction of a chemical potential weakens
the first order nature of the  transitions in our analysis.
This matches with results found in QCD on the lattice.
The weakening of the first order phase transition is demonstrated
for the chiral transition in the light quark mass regime
~\cite{deForcrand:2008vr,deForcrand:2006pv},
and is shown for the deconfinement transition
in the heavy quark mass regime~\cite{Langelage:2009jb,Kim:2005ck}.

{\bf Acknowledgements:} NE and KK are grateful for the support of
an STFC rolling grant. KK would like to thank Owe Philipsen, Sang-Jin Sin, Yunseok Seo, Johanna Erdmenger, and  Ingo Kirsch for discussions.
AG and MM are grateful for University of
Southampton Mayflower Scholarships.

\end{document}